\documentclass[a4paper,twocolumn,amsmath,amssymb,longbibliography]{revtex4-1}
\usepackage[english]{babel}
\usepackage{graphicx,color,soul}
\usepackage[binary-units=true]{siunitx}
\usepackage{multirow}

\newcommand{\al}{\alpha}

\newcommand{\bucky}{C$_{60}$}

\begin{document}

\title{Modeling epitaxial film growth of C$_{60}$ revisited}

\author{William Janke}
\author{Thomas Speck}
\affiliation{Institut f\"ur Physik, Johannes Gutenberg-Universit\"at Mainz, Staudingerweg 7-9, 55128 Mainz, Germany}
\email[Corresponding author: ]{thomas.speck@uni-mainz.de}

\begin{abstract}
  Epitaxial films evolve on time and length scales that are inaccessible to atomistic computer simulation methods like molecular dynamics (MD). To numerically predict properties for such systems, a common strategy is to employ kinetic Monte Carlo (KMC) simulations, for which one needs to know the transition rates of the involved elementary steps. The main challenge is thus to formulate a consistent model for the set of transition rates and to determine its parameters. Here we revisit a well-studied model system, the epitaxial film growth of the fullerene C$_{60}$ on an ordered C$_{60}$ substrate(111). We implement a systematic multiscale approach in which we determine transition rates through MD simulations of specifically designed initial configurations. These rates follow Arrhenius' law, from which we extract energy barriers and attempt rates. We discuss the issue of detailed balance for the resulting rates. Finally, we study the morphology of subatomic and multilayer film growth and compare simulation results to experiments. Our model enables further studies on multi-layer growth processes of C$_{60}$ on other substrates.
\end{abstract}

\maketitle


\section{Introduction}

Layer and cluster growth processes of organic molecules on dielectric or metallic surfaces have gained popularity in science and technology in recent years as a possible gateway to the manufacturing of new electronic structures at the molecular level~\cite{Flood04}. As Moore's law for today's silicon-based technology is predicted to flatten out in the near future, technological innovations will be necessary to further improve future electronic devices~\cite{SchulzEndofSilc,Ning2010SiliCMOSTech}. Another area is organic photovoltaics~\cite{graetzel12}. The self-assembly of nanostructures on substrates is a promising strategy~\cite{Barth05,Barth07,kuhnle09,lit15} and requires the understanding and control of growth processes to achieve target aggregate morphologies.

In particular carbon-based architectures such as the fullerene C$_{60}$ have been studied extensively. Morphologies and kinetics of C$_{60}$ film growth has been investigated experimentally on different substrates: graphite~\cite{LiuC60Exp}, pentacene~\cite{conrad09}, calcium fluoride~\cite{LoskePaper,Korn11}, mica~\cite{bomm14}, and iron~\cite{Picone16}. To access the large time and length scales required to predict morphologies in computer simulations requires coarse-graining. The dynamics of organic molecules on a lattice can be modeled as discrete jumps on time scales much longer than those of molecular vibrations~\cite{einax13}. The effective dynamics is necessarily stochastic and determined by a relatively small number of elementary discrete events. Such a dynamics is simulated efficiently employing the kinetic Monte Carlo (KMC) algorithm~\cite{KMCVoter2007} (also called Gillespie algorithm~\cite{Gillespie76,Gillespie77}). In principle, it requires the knowledge of the rates for all possible events, the determination of which for complex systems is a formidable task if not impossible. For the film growth of C$_{60}$, KMC simulations with a simplified rate catalogue have been conducted to reproduce experimental observations~\cite{LiuC60Sim,Korn11,bomm14,Kleppmann15,acevedo16,Kleppmann17}.

The number of possible transitions depends on the choice for the lattice and whether to neglect the distinction between A and B edge steps, which strongly influences cluster shapes~\cite{einax13}. One approach is to assume an Arrhenius law, which requires an attempt rate and an energy barrier for each possible transition. For C$_{60}$, energy barriers have been calculated from density functional theory (DFT)~\cite{GooseDFTDiffCalc} and molecular mechanics~\cite{Cantrell12}.

Setting up a KMC simulation, there is no unique way to determine the involved transition rates. One common approach is to describe the energy barriers and attempt rates through a simple self-consistent model involving a bond counting approach for the energy barriers with very few free parameters that are then either taken from literature or tuned to reproduce some experimental result~\cite{Korn11,bomm14}. However, without independent verification of the validity of those free parameters, the resulting transition rates might be unphysical, or the parameters can lose their intended interpretation. Another approach is to measure energy barriers and attempt rates for the transitions of interest directly, \emph{e.g.}, in MD simulations~\cite{LiuC60Sim}. However, as systematic or statistical uncertainties may be involved in those measurements, KMC models that make direct use of such transition rates are likely to be thermodynamically inconsistent (\emph{i.e.}, the rates break detailed balance).

\begin{figure}[t!]
  \includegraphics[width=3.2in]{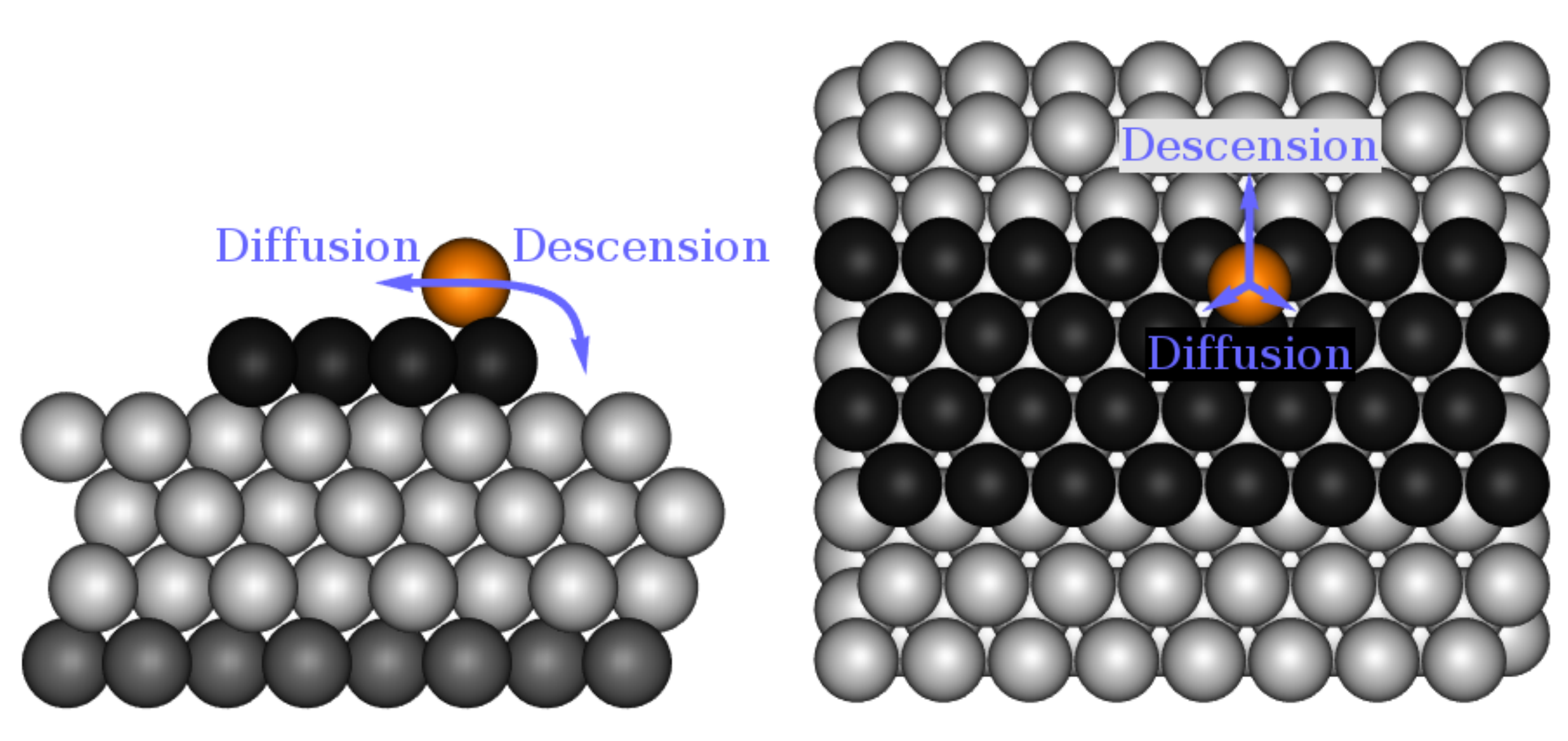}
  \caption{Example MD simulation setup. Right: Top view. Left: Side view. The dark gray particles are fixed in place to ensure an fcc crystal structure with a (111) surface. The three layers of light gray particles are freely evolving during the NVE simulation, but are under the effect of a Langevin thermostat or velocity rescaling at the beginning of the simulation. The black particles are set in a stable configuration where they are unlikely to move and set up the environment for the tagged orange particle, which is set in a state where the transitions of interest can be observed. Both black and orange particles are never under the effect of a thermostat or velocity rescaling. In this example setup, the transitions of interest are the descension and diffusion of the orange particle.}
  \label{fig:MDSimEx}
\end{figure}

\begin{figure*}[t]
  \includegraphics[width=\textwidth]{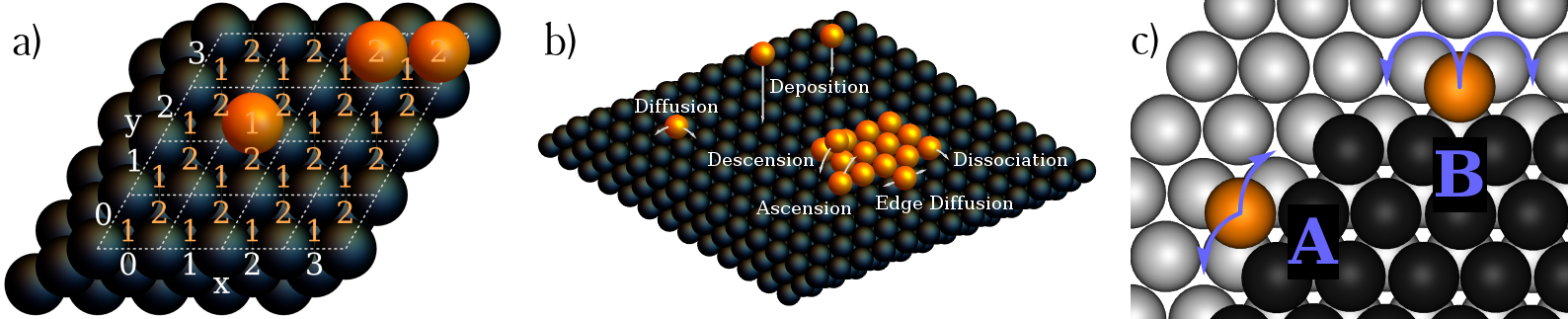}
  \caption{Geometry and involved transition types. (a)~Visualization of the sublattices. The unit cell is a parallelogram with two sublattice positions. A freely diffusing particle always jumps from one sublattice to the other, and particles that belong to the same cluster always occupy the same sublattice. Periodic boundary conditions are applied to $x$ and $y$ direction, while the $z$ direction consists of two layers. (b)~Visualization of the transition types in the simulated system. (c)~Difference between A and B step edge diffusion. While on A step edges, the orientation of the base layer facilitates the edge diffusion transition, on B step edges it does not. As a result, dissociation and reassociation can be the more probable trajectory at B step edges.}
  \label{fig:SystemVisualisations}
\end{figure*}

Here, we follow a hybrid approach and systematically determine the rates for the deposition and epitaxial growth of C$_{60}$ on a C$_{60}$ (111) surface from off-lattice, coarse-grained molecular dynamics simulations (Fig.~\ref{fig:MDSimEx}). We then construct a thermodynamically consistent model for the transition rates that is based on the MD simulation results. Establishing a thorough understanding of this system is crucial as it can then be applied to second layer processes in simulations with other substrates. The outline is as follows: We start by setting up MD simulations of interacting C$_{60}$ molecules to determine attempt rates and energy barriers for elementary transitions involving only a single C$_{60}$ molecule. To enforce the detailed balance condition, we will then reduce the number of parameters yielding a simplified model with 7 parameters, and compare the behavior of this model to the results obtained using the raw MD parameters. 


\section{Methods}

\subsection{MD Simulations: C$_{60}$ on C$_{60}$(111)}

To determine energy barriers and attempt rates for the diffusion of \bucky\ on a \bucky (111) surface, we set up MD simulations using LAMMPS~\cite{LAMMPSPlimpton}. The interaction between the individual \bucky\ molecules is modeled by a classical Girifalco pair potential~\cite{GirifalcoPot92,GirifalcoPot}
\begin{multline}
  u(s) = -\alpha\left(\frac{1}{s(s-1)^3}+\frac{1}{s(s+1)^3}-\frac{2}{s^4}\right) \\
  +\beta\left(\frac{1}{s(s-1)^9}+\frac{1}{s(s+1)^9}-\frac{2}{s^{10}}\right)
\end{multline}
with parameters $\alpha = \SI{46.7e-3}{\electronvolt}$ and $\beta = \SI{84.5e-6}{\electronvolt}$. Here, $s=r/R$ is the center-to-center distance $r$ of the two interacting molecules scaled by the nucleus-to-nucleus diameter of a C$_{60}$ molecule, $R=\SI{0.71}{\nano\metre}$. This potential yields a potential minimum of $E_G=\SI{277}{\milli\electronvolt}$ at a center-center distance of $r_\text{min}=\SI{1.005}{\nano\meter}$. Precision measurements show a very good agreement of this potential with experimental forces~\cite{ChiutuPotMeasure}. Still, one has to keep in mind that this coarse-grained potential removes rotational degrees of freedom as well as vibrations/deformations from the C$_{60}$ molecules, which can potentially play a role in the transition paths.

The C$_{60}$ substrate is modeled by four layers of which the bottom layer is immobilized. On top of the substrate, a stable configuration of deposited particles is set up with one or a few tagged particles left in metastable states. The life time of these metastable states as well as the frequency of outgoing transition types are observed to determine the associated transition rates. An example setup of such an MD simulation is shown in Fig.~\ref{fig:MDSimEx}. To initialize the system to a random starting configuration at a given temperature, a combination of the Langevin thermostat and velocity rescaling is applied. After equilibration, distributions of life times are obtained in NVE simulations.

\subsection{KMC Simulations}
  
In order to simulate at the length and timescales necessary for the observation of epitaxial growth processes, we employ a kinetic Monte Carlo (KMC) simulation. We implement two superimposed triangular sublattices with a lattice constant of $a=\SI{1}{\nano\metre}$ corresponding to the van der Waals diameter of C$_{60}$~\cite{khlob04}. A second layer with the same geometry can also be occupied as soon as a cluster has formed that can support it. A visualization of the lattice geometry can be found in Fig.~\ref{fig:SystemVisualisations}a. The basic types of implemented transitions are shown in Fig.~\ref{fig:SystemVisualisations}b.

The deposition of particles is part of the KMC rate catalogue and occurs randomly with a deposition rate $k_\text{Dep}=FA$, where $A$ is the system area in $\SI{}{\nano\metre^2}$. The molecular flux $F$ is set to a constant value of $\SI{5e-4}{\nano\metre^{-2}\second^{-1}}$. The $x$ and $y$ coordinates of deposition are picked randomly, and deposition can also occur directly onto the second layer of a cluster. If the target location is not valid (because the first layer is already taken and the second layer is either also taken or not sufficiently supported by particles below) a random walk is initiated at the target location to find a valid location for deposition in the proximity. Particles with more than four lateral neighbors are considered immobile in our simulation, while particles with one to four lateral neighbors have the possibility to move either by edge diffusion, ascension, descension, or dissociation. Furthermore, there are two types of edge diffusion transitions with different transition rates that have to be distinguished: edge diffusion along A step and along B step edges (see Fig.~\ref{fig:SystemVisualisations}c for a visualization of the distinction between the two). The transition rates of all these possible transitions have to be modeled and the model parameters will be determined using the data from MD simulations.


\section{Results}

\subsection{MD Simulations}

\begin{figure}[b!]
  \includegraphics[width=3.2in]{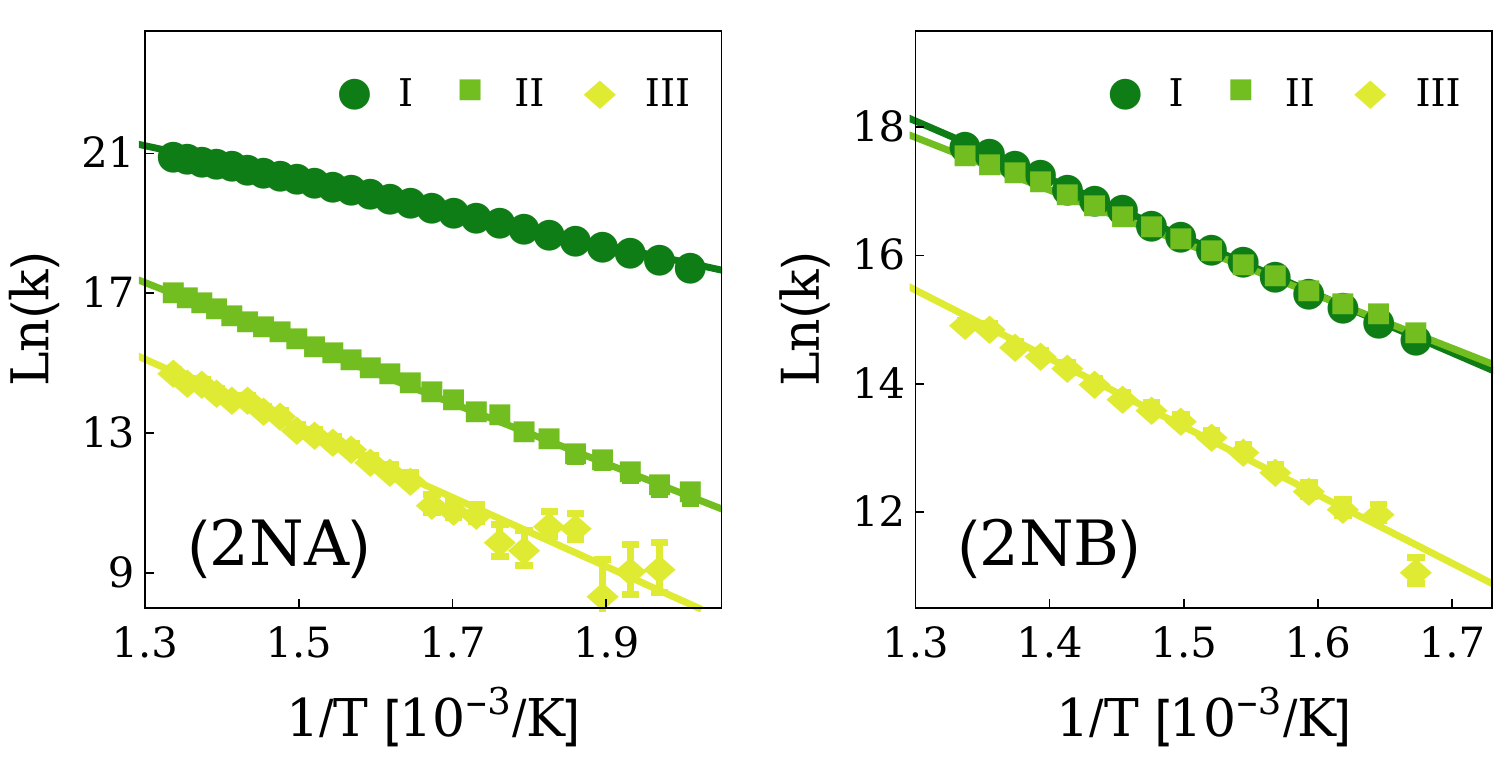}
  \caption{Example Arrhenius plots. The temperature range is varied depending on the configuration in order to observe as many transitions as possible. Left: Configuration (2NB) of table~\ref{tab:MDResultSummary} with a temperature range of $T\in [600,750]$K, sampling a total of $\approx 120000$ trajectories. Right: For configuration (2NA) with a temperature range of $T\in [500,750]$K, sampling a total of $\approx 2300000$ trajectories. Because in configuration (2NA) transition I had a much higher transition rate than the transitions II and III, a much larger amount of trajectories was needed to sufficiently sample all transitions.}
\label{fig:ExArrheniusPlots}
\end{figure}

We ran many simulations at several different temperatures $T$ in the range between $\SI{200}{\kelvin}$ up to $\SI{850}{\kelvin}$ and measure the total time the tagged particle has spent in the initial state at each temperature, $t_\text{tot}(T)$, as well as the number of occurrences of each transition of interest, $n_\al(T)$. The transition rates for every transition $\al$ are estimated through
\begin{equation}
  k_\al(T) \approx \frac{n_\al(T)}{t_\text{tot}(T)}.
\end{equation}
The transition rates $k_\al(T)$ are then fitted with an Arrhenius law to obtain an energy barrier $\Delta E_\al$ and an attempt rate $\nu_\al$ for the transition $\al$. The configurations investigated are visualized in Table~\ref{tab:MDResultSummary} along with the possible transition targets as well as their corresponding energy barriers and attempt rates, which are obtained from Arrhenius plots. Example Arrhenius plots for two of the configurations are given in Fig.~\ref{fig:ExArrheniusPlots}. In the following, when one of the transitions of Table~\ref{tab:MDResultSummary} is referred to, the reference will be comprised of the label on the image and the roman number of the transition target (\emph{e.g.} (F)I for the free diffusion transition).

\begin{table}
\centering
\begin{tabular}{c|c|c|cc}
Configuration & T & Jump to & $\Delta E [\SI{}{\milli\electronvolt}]$ & $\nu_0 [\SI{e12}{\hertz}]$ \\ 
\hline \\ [-1.5ex]
\multirow{4}{*}{\includegraphics[width=0.12\textwidth]{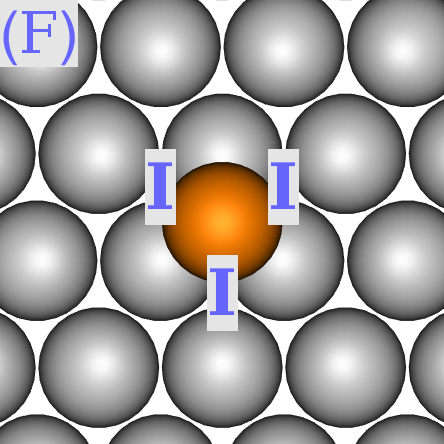}} & $200$ & I & $195.6(20)$ & $0.43(4)$ \\[6pt]
& $-$ &  &  & \\[6pt]
& $380$ & &  &  \\[6pt]
& $\SI{}{\kelvin}$ &  &  &  \\[6pt]
\hline \\ [-1.5ex]
\multirow{4}{*}{\includegraphics[width=0.12\textwidth]{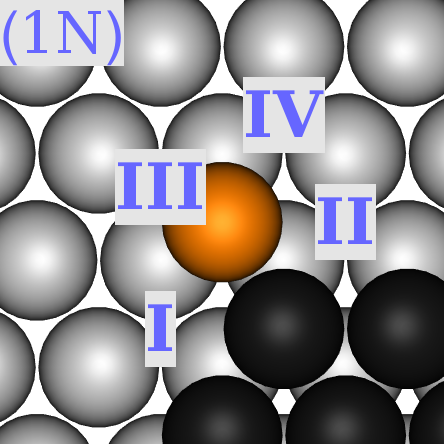}} & $350$ & I & $501(4)$ & $0.53(6)$ \\[6pt]
& $-$ & II &$188.3(6)$ & $0.262(4)$ \\[6pt]
& $550$ & III &$409(2)$ & $0.38(2)$ \\[6pt]
& $\SI{}{\kelvin}$ & IV &$458(4)$ & $0.25(4)$ \\[6pt]
\hline \\ [-1.5ex]
\multirow{4}{*}{\includegraphics[width=0.12\textwidth]{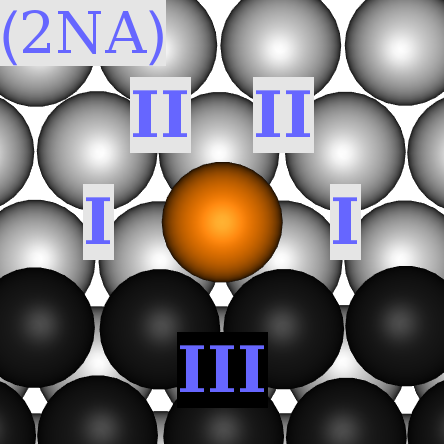}} & $500$ & I & $439(2)$ & $0.74(3)$ \\[6pt]
& $-$ & II &$744(5)$ & $1.23(12)$ \\[6pt]
& $750$ & III &$879(34)$ & $2.29(144)$ \\[6pt]
& $\SI{}{\kelvin}$ & &  &  \\[6pt]
\hline \\ [-1.5ex]
\multirow{4}{*}{\includegraphics[width=0.12\textwidth]{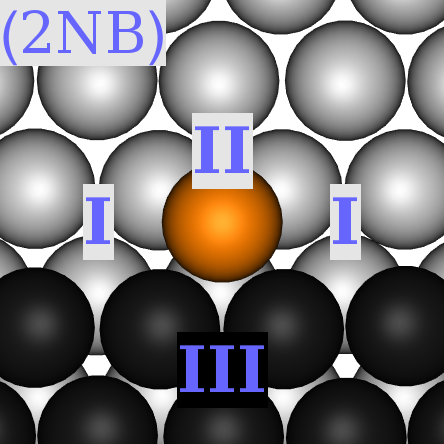}} & $600$ & I & $779(5)$ & $4.60(38)$ \\[6pt]
& $-$ & II &$708(4)$ & $2.46(18)$ \\[6pt]
& $750$ & III &$917(29)$ & $5.33(266)$ \\[6pt]
& $\SI{}{\kelvin}$ & &  &  \\[6pt]
\hline \\ [-1.5ex]
\multirow{4}{*}{\includegraphics[width=0.12\textwidth]{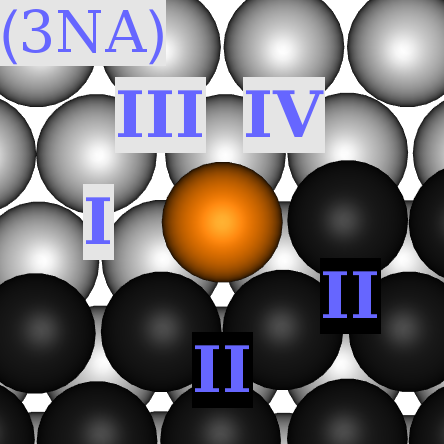}} & $680$ & I & $686(5)$ & $1.33(10)$ \\[6pt]
& $-$ & II & $1119(30)$ & $4.32(210)$ \\[6pt]
& $770$ & III & $1037(20)$ & $4.55(148)$ \\[6pt]
& $\SI{}{\kelvin}$ & IV & $1062(24)$ & $6.29(165)$ \\[6pt]
\hline \\ [-1.5ex]
\multirow{4}{*}{\includegraphics[width=0.12\textwidth]{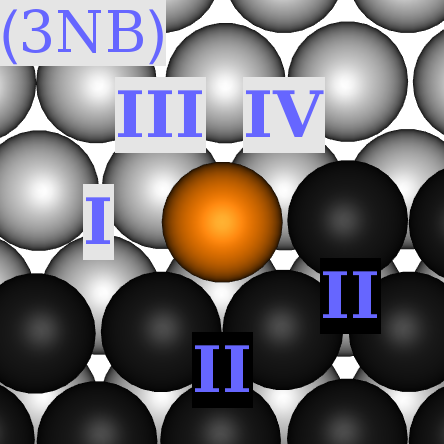}} & $680$ & I & $989(17)$ & $3.01(82)$ \\[6pt]
& $-$ & II & $1119(24)$ & $4.76(186)$ \\[6pt]
& $770$ &III & $967(13)$ & $3.49(73)$ \\[6pt]
& $\SI{}{\kelvin}$ & IV & $618(9)$ & $0.20(3)$  \\[6pt]
\hline \\ [-1.5ex]
\multirow{4}{*}{\includegraphics[width=0.12\textwidth]{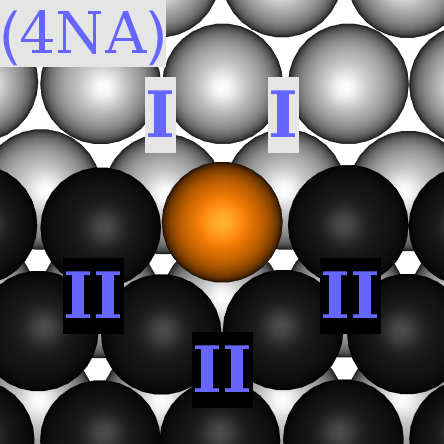}} & $750$ & I & $906(12)$ & $0.91(16)$ \\[6pt]
& $-$ & II &$1295(46)$ & $2.47(169)$ \\[6pt]
& $840$ &  &  &  \\[6pt]
& $\SI{}{\kelvin}$ & &  &  \\[6pt]
\hline \\ [-1.5ex]
\multirow{4}{*}{\includegraphics[width=0.12\textwidth]{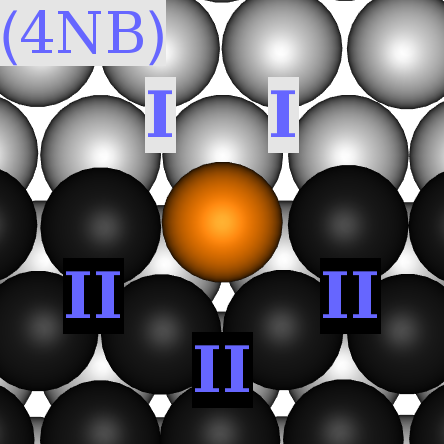}} & $750$ & I & $1270(18)$ & $10.4(28)$ \\[6pt]
& $-$ & II &$1311(39)$ & $3.83(217)$ \\[6pt]
& $840$ &  &  &  \\[6pt]
& $\SI{}{\kelvin}$ & &  &  \\[6pt]
\hline \\ [-1.5ex]
\multirow{4}{*}{\includegraphics[width=0.12\textwidth]{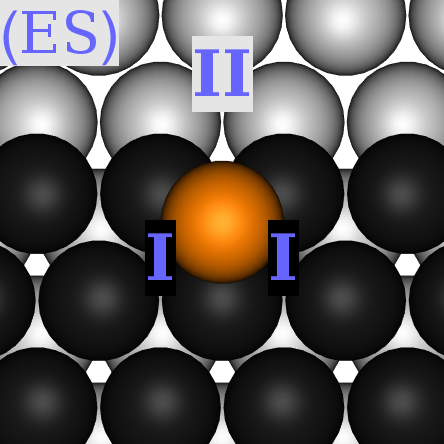}} & $250$ & I & $181.4(3)$ & $0.325(3)$ \\[6pt]
& $-$ & II & $300.2(13)$ & $0.385(18)$ \\[6pt]
& $400$ & &  &  \\[6pt]
& $\SI{}{\kelvin}$ & &  &  \\[6pt]
\hline
\end{tabular}
\caption{Summary of the results from our MD Simulations. The first column shows the configuration with tagged particle (orange) and the possible transitions (blue numerals).}
\label{tab:MDResultSummary}
\end{table}

\begin{table}[t]
  \begin{tabular}{c|c|c|c}
    Transition & our work & MM~\cite{Cantrell12} & MD~\cite{LiuC60Sim} \\
    \hline
    (F)I & 195.6(20) & 205(22) & 178(4) \\
    (1N)III & 409(2) & 448(25) & 429(57) \\
    (2NB)II & 708(4) & 717(29) & - \\
    (ES)II & 300.2(13) & 334(20) & - \\
    \hline
  \end{tabular}
  \caption{Comparison of energy barriers $\Delta E$ (in $\SI{}{\milli\electronvolt}$) with previous results from molecular mechanics (MM) and molecular dynamics (MD). Note that the MM results are consistently larger.}
  \label{tab:comp}
\end{table}

In comparison to previous estimates we find that our value for the free diffusion barrier of $E_D=\SI{195.6(20)}{\milli\electronvolt}$ falls between the values obtained previously: Gravil \emph{et al.}~\cite{GravilPPotDiffCalc} ($\SI{168}{\milli\electronvolt}$, pair potential calculations), Liu \emph{et al.}~\cite{LiuC60Sim} ($\SI{178(4)}{\milli\electronvolt}$, MD simulations), Cantrell and Clancy~\cite{Cantrell12} ($\SI{205(22)}{\milli\electronvolt}$, molecular mechanics), and Goose \emph{et al.}~\cite{GooseDFTDiffCalc} ($\SI{207}{\milli\electronvolt}$, DFT calculations). In Table~\ref{tab:comp}, we compare our values for energy barriers to available previous results. To get an estimate for the Ehrlich-Schw\"obel barrier of this system, we have to subtract the free diffusion barrier from the energy barrier for descension given by (ES)II,
\begin{multline}
  E_{ES} = E_{DES} - E_{D} \\ = \SI{300.2}{\milli\electronvolt}-\SI{195.6}{\milli\electronvolt} = \SI{104.6(24)}{\milli\electronvolt},
\end{multline}
which is in very good agreement with the estimate by Goose~\cite{GooseDFTDiffCalc} of $\SI{104}{\milli\electronvolt}$.

\subsection{Modeling the rates}

All rates measured in the MD simulations conform with the Arrhenius law so that we split transition rates $k=\nu e^{-\Delta E/k_\text{B}T}$ into attempt rate $\nu$ and energy barrier $\Delta E$, both of which are independent of temperature. Since the interaction between C$_{60}$ molecules is short-ranged (see appendix~\ref{sec:bondCount}), we simplify rates by assuming that they only depend on the initial state and the type $i$ of transition but not on the target state. We thus arrive at the parametrization
\begin{equation}
  \label{eq:rate}
  k_i(n,T) = \nu_i(n) e^{-\Delta E_i(n)/k_\text{B}T}
\end{equation}  
characterizing the initial state by the number $n$ of neighbors in the same layer. The possible types $i$ are $D$ (free diffusion and dissociation), $EDA$ (edge diffusion along A steps), $EDB$ (edge diffusion along B steps), $A$ (ascension), and $DES$ (descension).

\begin{figure}[b!]
  \includegraphics[width=3.2in]{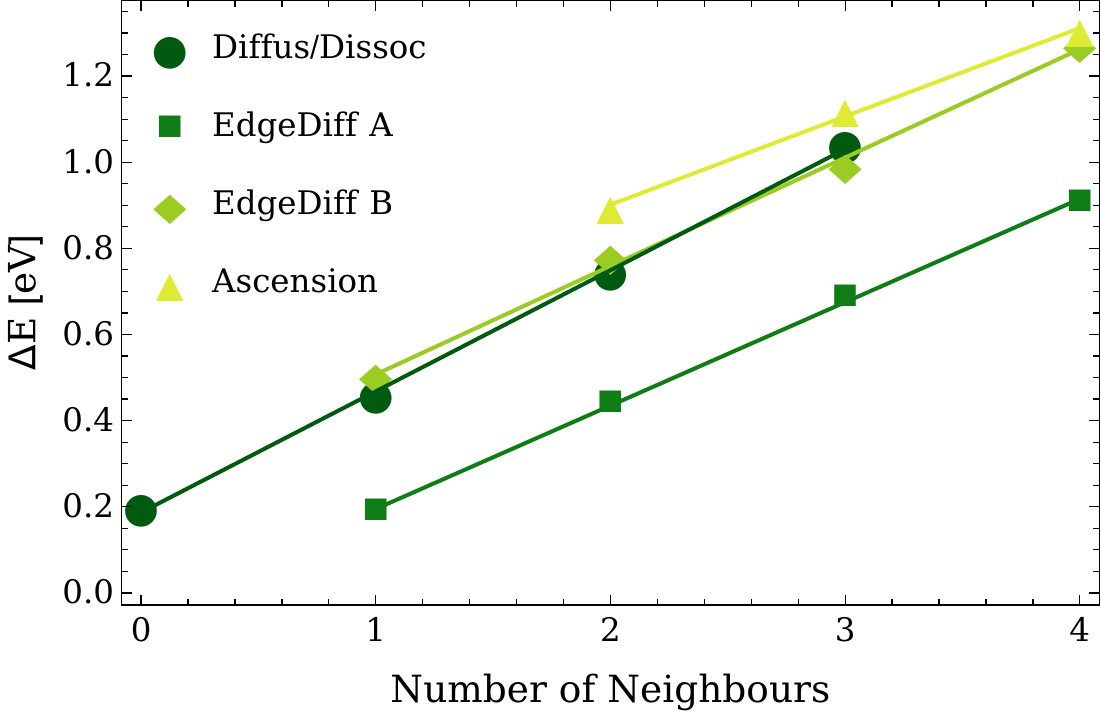}
  \caption{Energy barriers of the different transition types plotted against the number of initial neighbors. The dissociation barriers are taken from  (1N)IV, (2NA)II and (3NA)III, which are the longer dissociation paths compared to the ones in B step directions and are therefore the energy barriers for a complete dissociation. For the energy barriers of ascension there is no significant difference between the A step and B step case, so we took the average of the two respective measurements for this plot. The data for edge diffusion along an A step edge is comprised of (1N)II, (2NA)I, (3NA)I and (4NA)I, while edge diffusion along a B step edge is given by (1N)I, (2NB)I,(3NB)I and (4NB)I. All energy barriers are well described by linear fits (lines).}
  \label{fig:AllEnergyBarriers}
\end{figure}

In Fig.~\ref{fig:AllEnergyBarriers}, the energy barriers extracted from the MD results are plotted as a function of initial neighbors. We observe a linear dependence, which motivates to further parametrize the energy barriers as
\begin{equation}
  \label{eq:E}
  \Delta E_i(n) = E_i + \hat{n}_i E_{B,i}.
\end{equation}
Here, $E_i$ is a base energy barrier and $E_{B,i}$ is an effective bond energy for transitions of type $i$. For dissociation and the descension to the lower layer, $\hat{n}_i=n$ is equal to the number of initial lateral neighbors $n$. However, for edge diffusion and ascension transitions lateral bonds can be sustained during the transition, which leads to $\hat{n}_i=n-1$ for edge diffusion and $\hat{n}_i=n-2$ for ascension.

From the linear fits in Fig.~\ref{fig:AllEnergyBarriers} we can extract the effective bond energies $E_{B,i}$ for the different transition types, which are listed in Table~\ref{tab:energies}. The Girifalco potential we use in the MD simulations has an energy minimum of $E_G=\SI{277}{\milli\electronvolt}$, which is what we would expect as en energy barrier whenever a bond has to be broken completely to go through a transition. For the case of dissociation, the effective bond strength is in very good agreement with this assumption as $E_G$ is very close to our fit value of $E_{B,D}$. The effective bond strengths of edge diffusion along A and B step edges overlap within their margins of error and are both significantly lower than the full bond energy $E_G$. This was also expected as the transitional states of the edge diffusion transitions are mostly still close enough to the initial neighbors such that the transitioning molecule does not have to overcome the total bond energy in order to slide down into the target state. This implies that the assumption of short-ranged interaction and transition rates that only depend on the nearest neighbor configuration is not entirely valid and has to be seen as an approximation in our KMC simulations. The significantly lower effective bond strength of the ascension transition of only $\SI{205(10)}{\milli\electronvolt}$ per bond was unexpected and at this point is not completely understood. An analysis of the exact transition paths of ascension may lead to interesting insights but is not performed in this work since the ascension transitions have a high enough energy barrier to not play a big role in the following KMC simulations. We have not set up an extra set of simulations to determine the effective bond strength $E_{B,DES}$ for descension transitions, so we assume that it is equal to $E_{B,D}$ as it also has a long transition path in which all initial bonds have to be overcome.

\begin{table}[t]
  \centering
  \begin{tabular}{rl|c|c}
    Transition type & $i$ & $E_{B,i}$ [$\SI{}{\milli\electronvolt}$] & $E_i$ [$\SI{}{\milli\electronvolt}$] \\
    \hline
    diff./dissociation & $D$ & 276(5) & 192(6) \\
    A edge step & $EDA$ & 245(4) & 190(5) \\
    B edge step & $EDB$ & 254(10) & 508(13) \\
    ascension & $A$ & 206(12) & 901(14) \\
    descension & $Des$ & 276(5) & 300(1) \\
    \hline
  \end{tabular}  
  \caption{Fitted effective bond energies $E_{B,i}$ and base energy barriers $E_i$, cf. Eq.~\eqref{eq:E}.}
  \label{tab:energies}
\end{table}

\begin{table}[b!]
  \centering
  \begin{tabular}{cl|c|c|c|c|c}
	  & &\multicolumn{5}{c}{$n$} \\
    Transition type & $i$ & 0 & 1 & 2 & 3 & 4 \\
    \hline
    diff./dissociation & $D$ & 0.43 & 0.25 & 1.23 & 4.02 & - \\
    A edge step & $EDA$ & - & 0.26 & 0.74 & 1.33 & 0.91 \\
    B edge step & $EDB$ & - & 0.53 & 4.60 & 3.01 & 10.4 \\
    ascension & $A$ & - & - & 3.81 & 4.54 &  3.15\\
    descension & $Des$ & 0.38 & 0.38 & 0.38 & 0.38 & 0.38 \\
    \hline
  \end{tabular}  
  \caption{Effective attempt rates $\nu_{i,n}$ as a function of neighbors $n$ used in the RawMD simulations (in units of $\SI{}{\tera\hertz}$).}
  \label{tab:attemptrates}
\end{table}

The base energy barriers $E_i$ obtained from the fits of Fig.~\ref{fig:AllEnergyBarriers} and from the MD simulation for descension are listed in Table~\ref{tab:energies}. The fact that edge diffusion along an A step edge is initialized by a transition path very similar to the free diffusion transition results in a base energy barrier $E_{EDA}$ that is close to the free diffusion barrier $E_D$. The base energy barrier for edge diffusion along B step edges is significantly larger than for the A step case, which is important for the formation of the triangular star shaped clusters that have been observed in experiments~\cite{LiuC60Exp,LoskePaper}. These star shapes evolve because of two effects that stem from the difference of A and B step barriers. First, the higher mobility at A step edges increases the chance of finding a state with high coordination, while particles on B step edges are likely to stick to an initial site with low coordination. Secondly, when particles reach a ``corner'' state [see (1N) and (3NA)/(3NB) in Tab. \ref{tab:MDResultSummary}] they are more likely to transition into the direction of an A step edge. As a result, trajectories from A step to B step edges are less likely to occur than from B step to A step edges. Since an ascending molecule has to overcome almost the full adsorption energy, the base energy barrier $E_A$ is almost equal to the adsorption energy of \bucky\ on \bucky (111), which is around $\SI{930}{\milli\electronvolt}$ (calculated with the Girifalco potential at zero Kelvin).

The attempt rates do not show any specific behavior in terms of the number of initial neighbors so we do not describe them with a model and just take the values as measured in the MD simulations. The exact values of the attempt rates $\nu_{i,n}$ are provided in Table~\ref{tab:attemptrates}.

In the following KMC simulations, we will consider two model variants. The first one (termed ``RawMD'' model) is given by the energies and attempt rates just described. In addition, we consider a further simplified version (the ``Simple'' model), where we only use one free attempt rate parameter $\nu_i(n)=\nu_0$. The second distinction between the two models is made with respect to the effective bond strengths $E_{B,i}$. In the ''RawMD'' model we allow for the transition types to have different effective bond strengths as we have measured slight differences in MD simulations. In the ''Simple'' model we discard these differences and demand that all transitions scale with the same effective bond strength $E_{B,i}=E_B$. In summary, for the ''RawMD'' model we end up with 23 parameters for attempt rates and 10 for energy barriers, while the ''Simple'' model only has one parameter for attempt rates and 6 for energy barriers.

\subsection{KMC Simulations}
\label{sec:KMCSims}

\subsubsection{Detailed balance}

One important aspect of extracting rates from MD simulations is their thermodynamic consistency. The atomistic dynamics is derived from a Hamiltonian and thus obeys detailed balance guaranteeing the absence of (steady) dissipation. For the KMC simulations, the detailed balance condition can be expressed as $p_{\mathcal C}k_{\mathcal C\to\mathcal C'}=p_{\mathcal C'}k_{\mathcal C'\to\mathcal C}$, where $\mathcal C$ is a configuration (the position of all molecules), $p_{\mathcal C}$ is the stationary probability to observe this configuration, and $k_{\mathcal C\to\mathcal C'}$ are the transition rates to go from one configuration to another. Possible transitions are $k_i(n)$, cf. Eq.~\eqref{eq:rate}. The question is whether these rates with the extracted parameters obey detailed balance.

\begin{figure}[t]
  \includegraphics[width=3.2in]{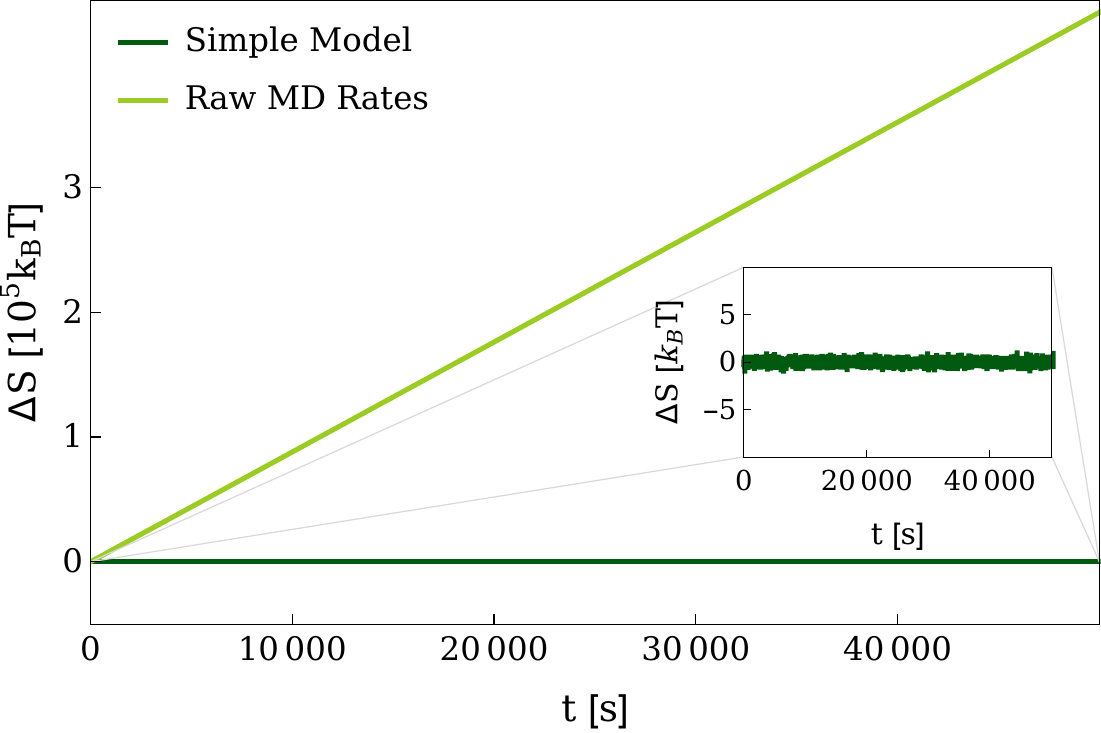}
  \caption{Comparison of the average entropy production between the RawMD and Simple model plotted against time $t$. The ensemble average is calculated from $100$ trajectories at a temperature of $T=\SI{318}{\kelvin}$.}
  \label{fig:EntropyProd}
\end{figure}

A simple and practical way to test this condition is to calculate the stochastic entropy production~\cite{Seifert2005EntropyProd}. To this end, we prepare a small $10\times 10$ unit cell system at a temperature of $T=\SI{318}{\kelvin}$ without deposition and with nine particles assembled into a cluster. During the simulation, the particles of this cluster will evolve through the different transition types in various successions and we can observe if any net entropy production occurs. After every KMC step, we calculate the change of (dimensionless) entropy
\begin{equation}
  \delta s = \ln\frac{k_i(n)}{k_j(m)}
  \label{eq:entropy}
\end{equation}
as the logarithm of the ratio between the rate $k_i(n)$ of the transition that occurred (the new number of neighbors is $m$) and the rate $k_j(m)$ of the transition that would reverse the KMC step. Summing up these single contributions along a trajectory yields the total entropy production $\Delta S$. After some initial relaxation, this total entropy production should fluctuate around zero if detailed balance were to be obeyed. As shown in Fig.~\ref{fig:EntropyProd}, this is not the case for the RawMD model, which exhibits a linear increase corresponding to a substantial dissipation of $8.81 k_\text{B}T$ of entropy per second. In contrast, our Simple model does obey detailed balance as demonstrated in Fig.~\ref{fig:EntropyProd}. This demonstrates that simply transferring rates from MD simulations is likely to break detailed balance and thus yield a thermodynamically inconsistent model, which can lead to severe artifacts.

\subsubsection{Morphologies}

\begin{figure}[h!]
  \includegraphics[width=3.2in]{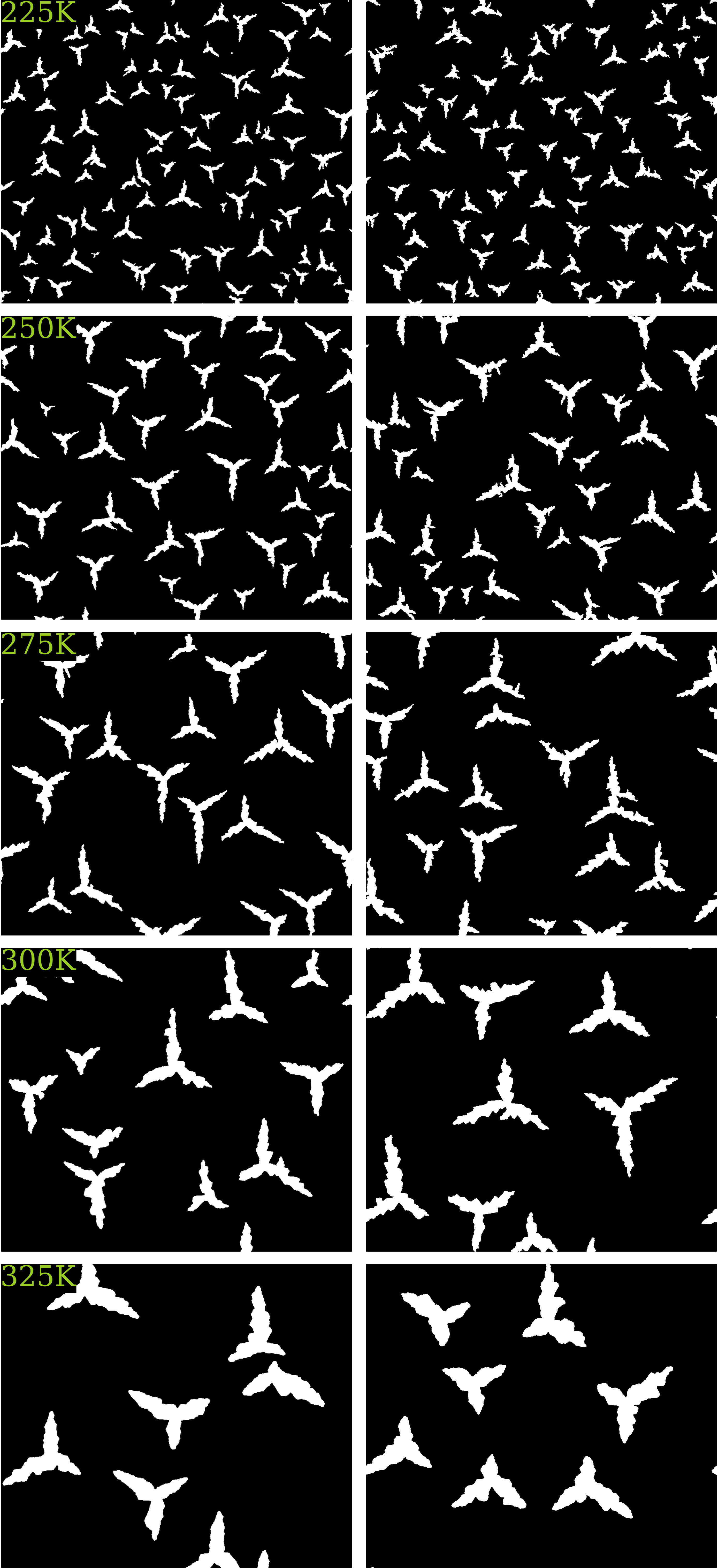} 
  \caption{KMC simulation snapshots of RawMD (left) and Simple (right) model at several temperatures (increasing from top to bottom). Parameters for the Simple model are $E_B=\SI{235}{\milli\electronvolt}$ and $\nu_0=\SI{0.25}{\tera\hertz}$. The snapshots show an area of $\SI{1000}{\nano\metre}\times \SI{866}{\nano\metre}$ and each contains $100000$ molecules corresponding to $10\%$ coverage.}
  \label{fig:MorphoSnapshots}
\end{figure}

\begin{figure}[h!]
  \includegraphics[width=3.2in]{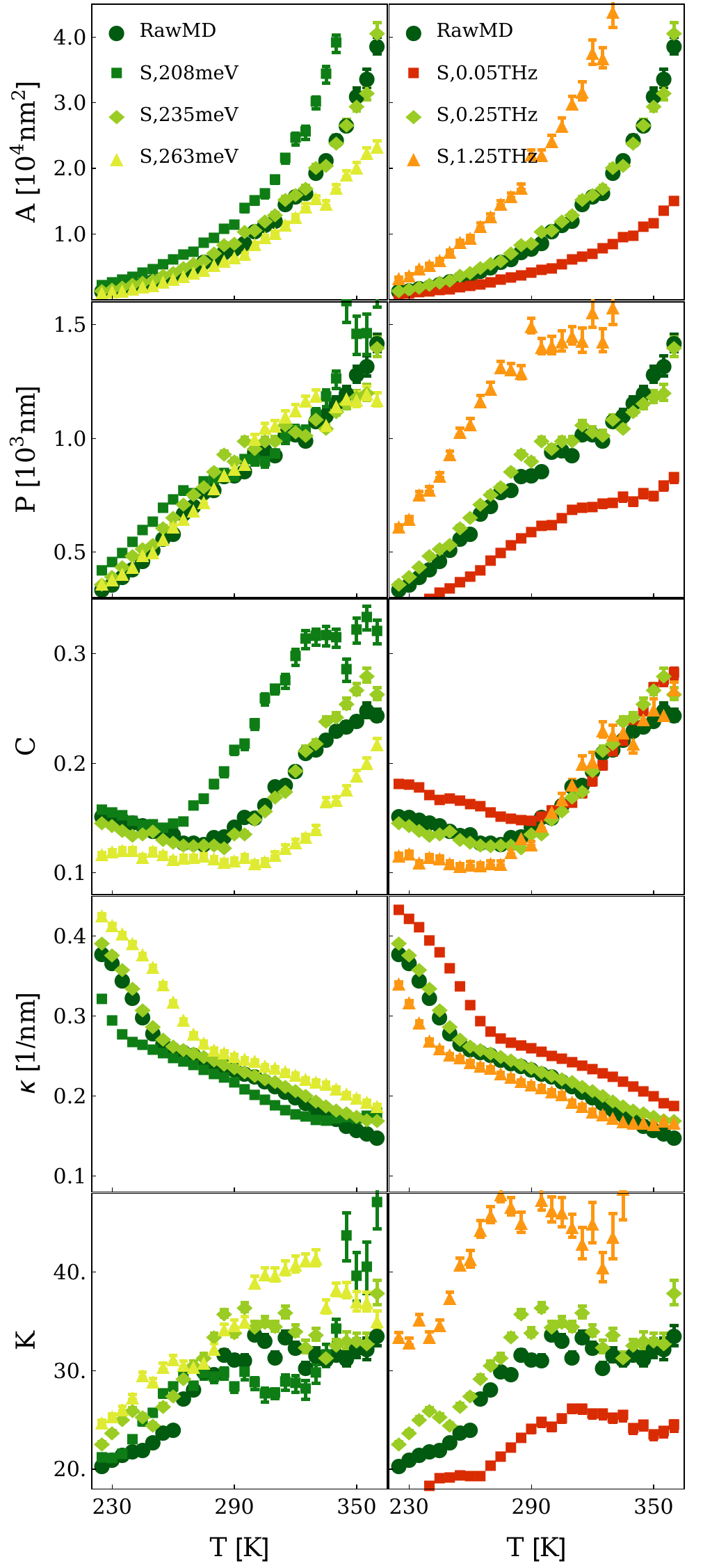}
  \caption{Average cluster features of the two KMC models in a temperature range from $\SI{225}{\kelvin}$ to $\SI{360}{\kelvin}$. Left: Comparison of the RawMD Model to three Simple models with varying effective bond strength $E_B$ and a constant attempt rate of $\nu_0=\SI{0.25}{\tera\hertz}$. Right: Comparison of the RawMD Model to three Simple models with varying attempt rate $\nu_0$ and a constant effective bond strength of $E_B=\SI{235}{\milli\electronvolt}$. As suggested by the shape descriptors, the resulting cluster morphologies are very similar.}
  \label{fig:ShapeDiscriptorPlots}
\end{figure}

To compare the emerging cluster morphologies of the two models, KMC simulations are conducted in a temperature range of $225-\SI{360}{\kelvin}$ with a system size of $A\approx \SI{0.866}{\micro\metre^2}$ ($1000$ times $1000$ unit cells of size $\SI{0.886}{\nano\metre^2}$). The system starts with an empty lattice and deposition of particles proceeds until a coverage of 10\% is reached. This deposition phase takes about $t_D\approx\SI{230}{\second}$ to complete. The simulation then continues without deposition to enable relaxation of the clusters for another $t_R\approx\SI{460}{\second}$. Figure~\ref{fig:MorphoSnapshots} shows representative snapshots for five different temperatures. With the bare eye, no significant differences between the snapshots of the RawMD and Simple model can be spotted.

To compare the morphologies of the different models more quantitatively, three geometric properties of the arising clusters are measured: The covered area $A$, the perimeter $P$ and the mean border curvature $\bar{\kappa}$. Area and perimeter are directly calculated from the number of molecules in the cluster $N$ and the number of molecules at its edge $N_E$ via
\begin{equation}
  A = N \times \SI{0.866}{\nano\metre^2}, \qquad
  P = N_E \times \SI{1}{\nano\metre}.
\end{equation}
Details for the calculation of $\bar\kappa$ are found in appendix~\ref{sec:curv}. We also determine two dimensionless quantities, the circularity $C$ and the dimensionless curvature $K$ defined through
\begin{equation}
  C = \frac{4\pi A}{P^2}, \qquad K=\frac{\bar{\kappa}P}{2\pi},
\end{equation}
which both return unity for circles of any size.

In Fig.~\ref{fig:ShapeDiscriptorPlots}, we plot the different shape descriptors as a function of temperature comparing the Simple model with the more elaborate RawMD model. We see that an excellent agreement of cluster morphologies can be achieved using an effective bond strength of $E_B=\SI{235}{\milli\electronvolt}$ and an attempt rate of $\nu_0=\SI{0.25}{\tera\hertz}$. The value of $E_B$ is closest to the effective bond strengths for edge diffusion in the RawMD model as these are the most important transitions when it comes to cluster relaxation. However, using this small value for $E_B$ the dissociation rate is increased in the Simple model, which would lead to a lower cluster density. To compensate for this, the attempt rate $\nu_0$ is lower than the attempt rate for free diffusion in the RawMD model, which increases the cluster density to produce the correct cluster sizes. Variation of the model parameters from their optimal values can lead to significant differences in the shape descriptors.

\subsection{Multilayer Growth}

\begin{figure}[b!]
  \includegraphics[width=3.2in]{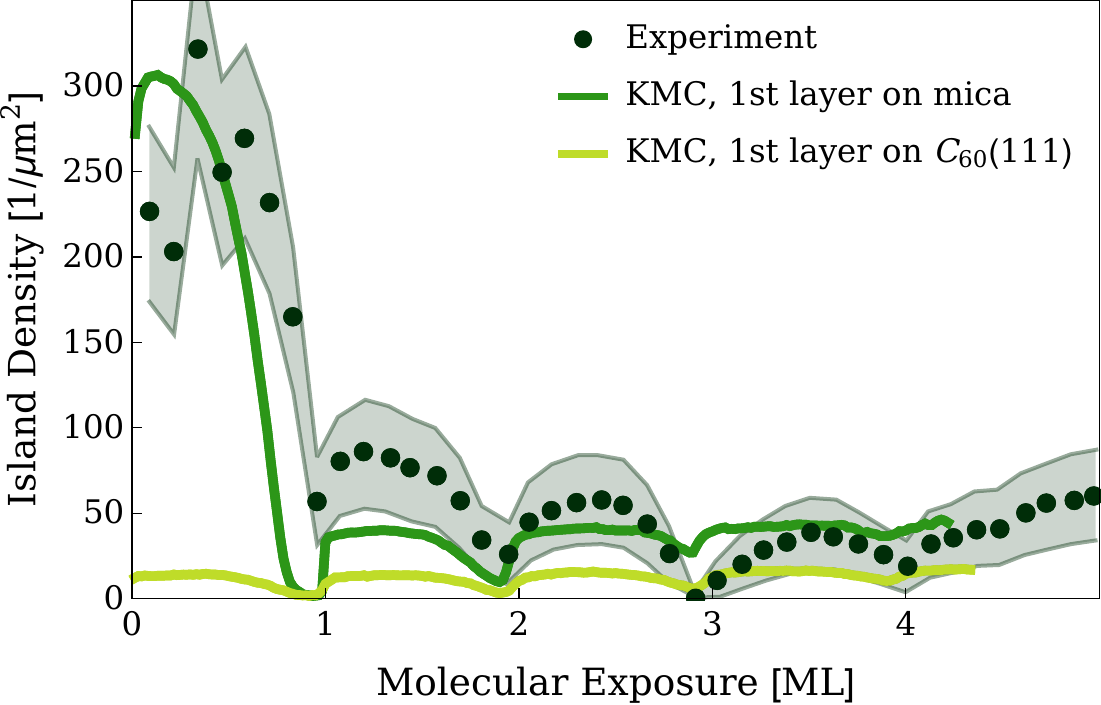}
  \caption{Cluster density as a function of molecular exposure (solid symbols) at $T=\SI{333}{\kelvin}$. Also plotted are results from the optimal Simple KMC model: unmodified model for growth on a clean C$_{60}$ substrate (bright line) and with modified rates to reproduce the cluster density in the first layer (dark line).}
  \label{fig:IslandDensEvolution}
\end{figure}

We now turn to the consecutive growth of multiple layers, for which experimental results are available. In particular, we consider the data published in Ref.~\citenum{bomm14}. One quantity that can be extracted from scattering data is the average cluster density~\cite{Renaud03}, which is plotted in Fig.~\ref{fig:IslandDensEvolution} as a function of the mean film height measured in monolayers (ML, $\text{molecular exposure}=F/\bar n\times\text{time}$ with $\bar n$ the density of a full monolayer). For the molecular flux $F$ we now chose the same value as in Ref.~\citenum{bomm14} corresponding to $0.1$ monolayers per minute (ML/min). Starting from a high value, the cluster density rapidly decays and shows oscillations with minima corresponding to the filling of a complete layer. Also shown are results from KMC simulations employing the optimal Simple model [cf. Sec.~\ref{sec:KMCSims}]. While the period of the oscillations is recovered, our simulations yield cluster densities that are consistently too small, with the discrepancy exceeding one order of magnitude in the first and second layer. Moreover, in contrast to the experimental data, the simulated peak density in each layer is roughly independent of time.

\begin{figure*}[t]
  \includegraphics[width=\textwidth]{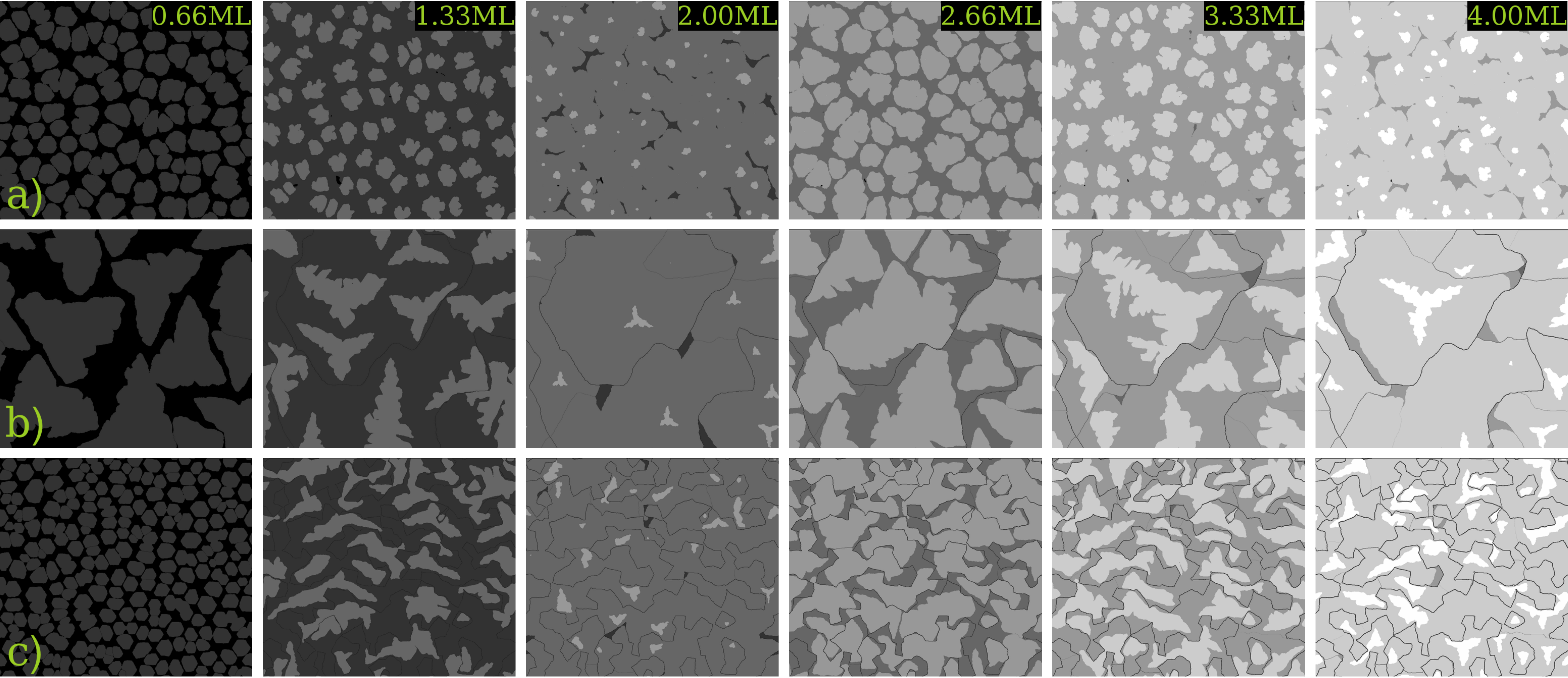}
  \caption{Snapshots at different stages of the multilayer growth simulations at a temperature of $T=\SI{333}{\kelvin}$ and a flux of $F=0.1\text{ML/min}$. (a)~Simplified KMC model from Ref.~\citenum{bomm14}. (b)~Simple model derived in Sec.~\ref{sec:KMCSims}. Note [grain boundaries] (c)~Using the Simple model for the second and higher layers. In the first layer, diffusion barrier and attempt rate have been altered to reproduce the experimental first layer cluster densities.}
  \label{fig:MultilayerSnaps}
\end{figure*}

In the same paper Ref.~\citenum{bomm14}, experimental results are reproduced numerically using a simplified KMC model (no sublattices, no distinction between A and B steps). The parameters of that model disagree with the parameters obtained here (and in other studies): $E_{D,\text{Bommel}}=\SI{540}{\milli\electronvolt}$ for the diffusion barrier compared to $E_D=\SI{192}{\milli\electronvolt}$ and $E_{B,\text{Bommel}}=\SI{130}{\milli\electronvolt}$ for the bond strength compared to $E_\text{B}=\SI{235}{\milli\electronvolt}$ (we recall that the bare Girifalco potential yields $E_G=\SI{277}{\milli\electronvolt}$). The reason for the different behaviors of cluster densities is revealed by comparing snapshots of the corresponding KMC simulations in Fig.~\ref{fig:MultilayerSnaps}: the simplified KMC simulations show many small clusters [Fig.~\ref{fig:MultilayerSnaps}a and Supplementary Video 1] while our model leads to the formation of a few large clusters [Fig.~\ref{fig:MultilayerSnaps}b and Supplementary Video 2]. Also the morphologies disagree, in the simplified model the clusters are round compared to triangular ``star'' shapes. This difference can mostly be attributed to neglecting the distinction between A/B type edge diffusion as well as the low value for the bond strength of $\SI{130}{\milli\electronvolt}$.

An import difference between the simulations and the experiments is that the later have been performed on mica (KAl$_3$Si$_3$O$_{12}$H$_2$) and not on a clean C$_{60}$(111) surface as assumed in our simulations, which may lead to impurities and ``grain boundaries''. Our KMC simulations include a mechanism that leads to such grain boundaries. Due to the use of two sublattices, different clusters do not merge in our model when initiated on different sublattices. This leads to grain boundaries between clusters, which persist in the following layers as observed in Fig.~\ref{fig:MultilayerSnaps}b and might explain the differences with the experiments: First, the diffusion barrier of C$_{60}$ is increased, which yields a much larger peak cluster density in the first layer. Second, mica has a lattice constant of $\SI{0.5}{\nano\metre}$, which is half that of the diameter of a C$_{60}$ molecule. Clusters on different effective sublattices cannot coalesce and thus create grain boundaries, which increase the peak cluster density also in the following layers.

To test this hypothesis, we have modified the dynamics for the first layer: no distinction between A and B type edge diffusion, an ascension barrier of $E_{A,\text{mica}}=\SI{1150}{\milli\electronvolt}$ chosen large enough to guarantee layer-by-layer growth, and, most importantly, the diffusion barrier and attempt rate read
\begin{equation}
  E_{D,\text{mica}} = \SI{655}{\milli\electronvolt}\qquad
  \nu_{0,\text{mica}} = \SI{4e15}{\hertz},
\end{equation}
which have been determined through a parameter tuning to reproduce the experimental cluster densities in the first layer (data from Ref.~\citenum{bomm15phd}). In Fig.~\ref{fig:MultilayerSnaps}c, we show the resulting temporal evolution of the film growth (see also Supplementary Video 3). It can be seen that the high cluster density of the first layer has a strong impact on the cluster densities of the subsequent layers because of the hindering effect of the grain boundaries on the diffusion process. Moreover, the time evolution of the cluster density is now in much better agreement with the experimental data at substrate temperature $T=\SI{333}{\kelvin}$ (Fig.~\ref{fig:IslandDensEvolution}, dark green line).

\begin{figure}[t]
  \includegraphics[width=3.2in]{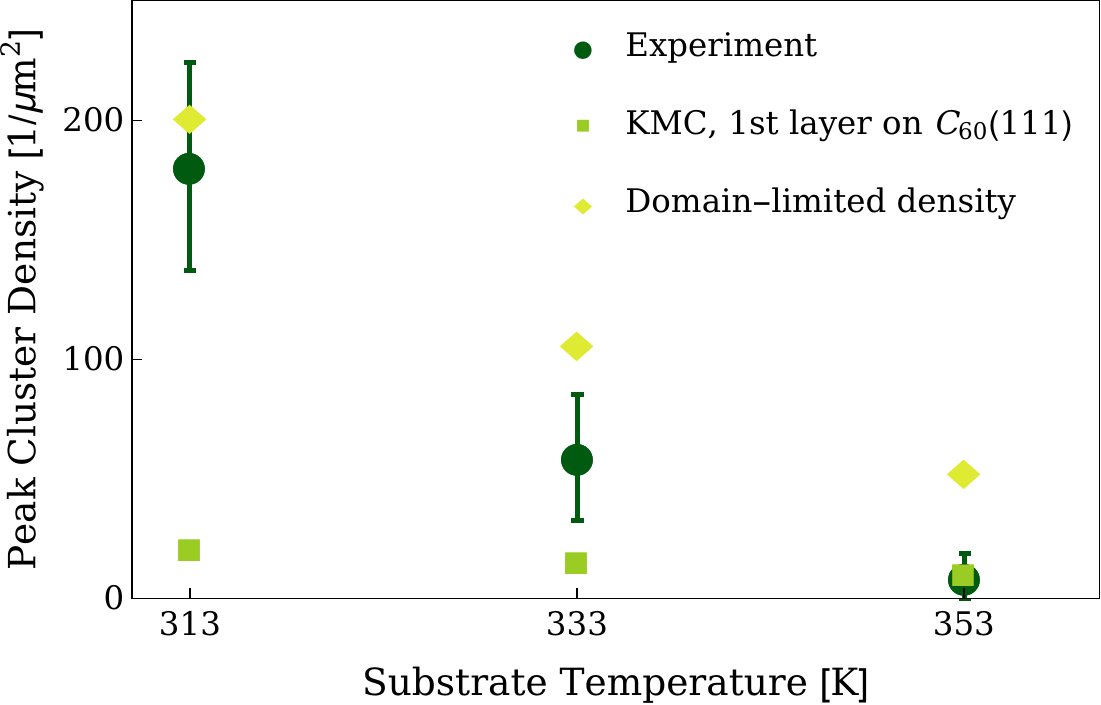}
  \caption{Peak cluster densities of the second C$_{60}$ layer at the three experimentally observed temperatures (discs). Also shown are the KMC prediction of the optimal Simple model (diamonds) and the upper bound Eq.~\eqref{eq:n_domains}.}
  \label{fig:MaxIslandDens}
\end{figure}

In Fig.~\ref{fig:MaxIslandDens}, we plot the peak cluster density of the second layer as a function of substrate temperature. Interestingly, the experiment perfectly matches the result of the simulations using our Simple model without a modified first layer dynamics at the higher temperature $T=\SI{353}{\kelvin}$. At the lower temperature $T=\SI{313}{\kelvin}$, our simulations predict a minute increase while the experiments show an even stronger increase of the peak cluster density. To estimate the effect of the grain boundaries on the peak cluster density in the second layer, we have numerically determined (on a hexagonal lattice) the number of connected domains of initial clusters with $m=(1/0.5)^2=4$ different species. We find
\begin{equation}
  \label{eq:n_domains}
  n_\text{domains} \approx 0.38 n_1,  
\end{equation}
where $n_1$ is the peak cluster density in the first layer. At the low temperature $T=\SI{313}{\kelvin}$, we find a good agreement of the experimental result with this prediction. This suggests that at low temperatures, cluster growth is indeed limited to domains enclosed by grain boundaries. At high temperatures, diffusion of C$_{60}$ molecules can overcome these barriers and we recover the behavior on a clean C$_{60}$ substrate.

\section{Conclusion}

We have studied the epitaxial film growth of C$_{60}$ molecules. Our central result is the systematic determination of a set of energy barriers (Table~\ref{tab:energies}) and attempt rates (Table~\ref{tab:attemptrates}) from many MD simulations of specific initial configurations. The bare rates do not obey detailed balance, which can be enforced through a single attempt rate $\nu_0$. A second simplification arises through employing a single effective bond energy $E_B$. We determine these two effective parameters through matching cluster morphologies, which yields
\begin{equation}
  E_B = \SI{235}{\milli\electronvolt}, \qquad \nu_0 = \SI{0.25}{\tera\hertz}.
\end{equation} 
We have then investigated multilayer growth. Differences to the experimental results of Ref.~\citenum{bomm14} are attributed to the mica substrate, the smaller lattice constant of which might induce grain boundaries between growing clusters which limits their size and increases their number. It would be interesting to directly resolve morphologies in the experiments.

While the C$_{60}$ on C$_{60}(111)$ system with the coarse-grained Girifalco potential is especially well suited, the method we have presented here to construct the rate parameters is generic and applicable to other combinations of organic molecule and substrate for which classical force fields are available. In particular, our results pave the way to investigate more systematically the molecular dewetting of organic molecules on non-metallic substrates~\cite{rahe12}. For more complex systems, the transition types and thus the number of parameters increases very quickly and might be prohibitively large to be determined from dedicated MD simulations. Another restriction on the part of the MD simulations is the feasibility of simulating the necessary timescales to observe the transitions of interest over a suitable range of temperatures. However, estimating rates can be achieved with less statistics and thus computational effort than spent here.

\begin{acknowledgments}
  We thank A. K\"uhnle for stimulating discussions. We acknowledge financial support by the Deutsche Forschungsgemeinschaft (grant no. 319880407). Numerical computations have been carried out on the MOGON II Cluster at ZDV Mainz.
\end{acknowledgments}

\appendix

\section{Bond Counting}
\label{sec:bondCount}

In KMC simulations of thin film growth it is common to use a bond counting approach [in our case Eq.~\eqref{eq:E}] to model energy barriers for transitions of particles that are in contact with a cluster.

\begin{figure}[t]
  \includegraphics[width=3.2in]{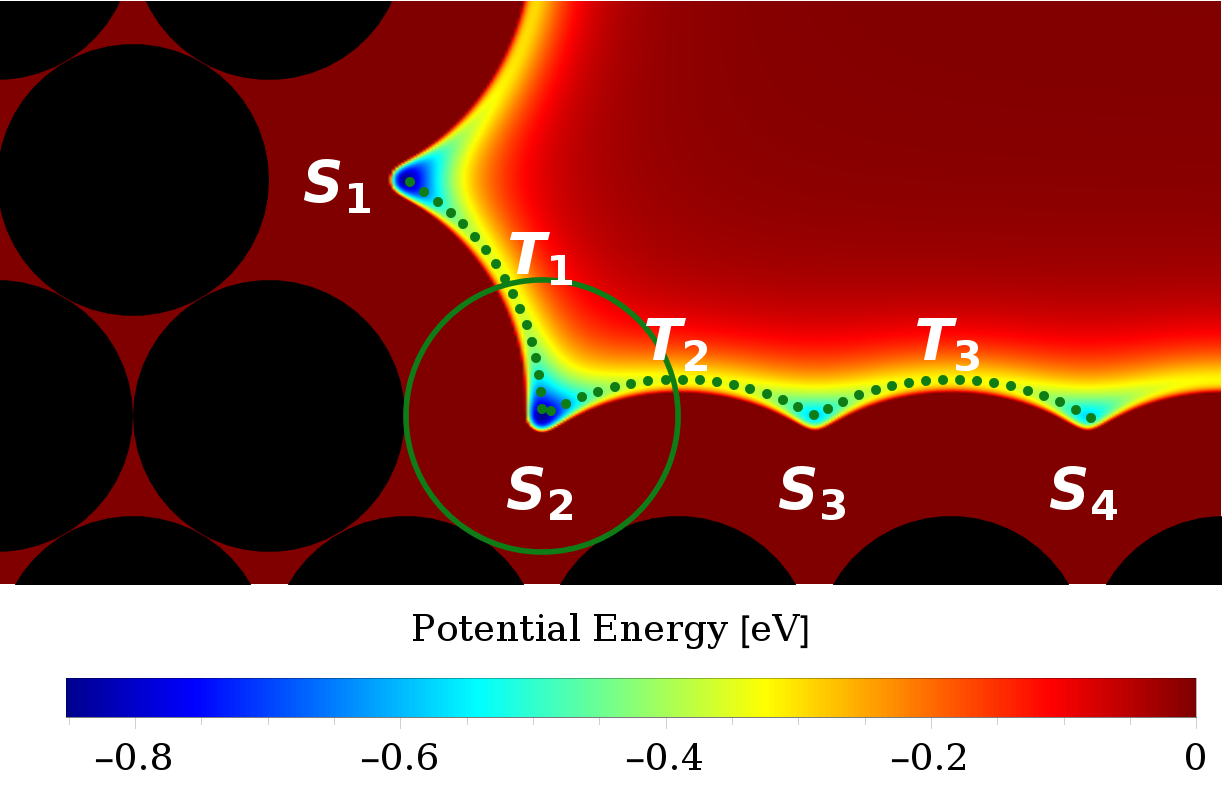}
  \includegraphics[width=3.2in]{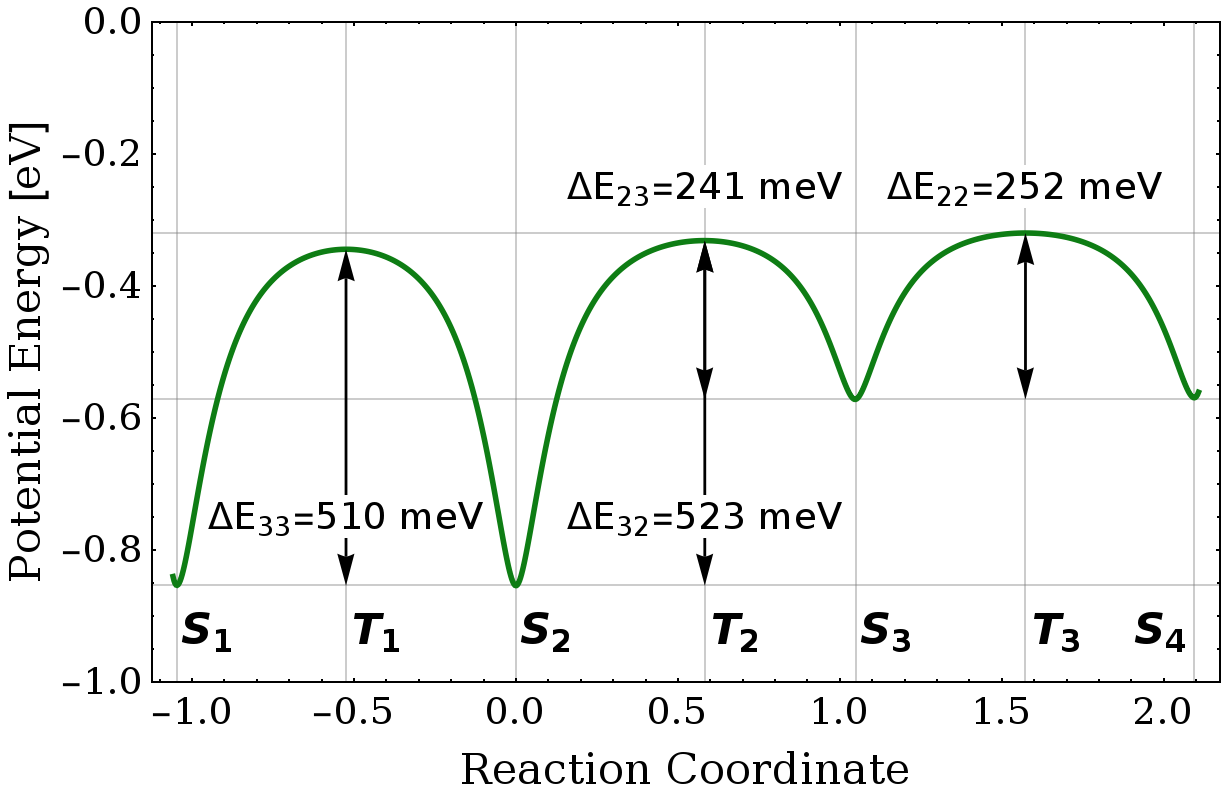}
  \caption{Top: Density plot of the potential energy of a sample molecule (dark green circle) in a given configuration of fixed cluster molecules (black disks, viewed from above) interacting via the Girifalco potential (the substrate is omitted). The dotted dark green line represents a possible reaction path connecting the four states $S_1,\dots,S_4$ passing through the transitional states $T_1,T_2$ and $T_3$. Bottom: Plot of the potential energy along the dotted reaction path. The energy barriers $\Delta E_{ij}$ ($i$ the number of initial neighbors, $j$ the number of final neighbors) are only weakly affected by $j$.}
  \label{fig:BondCounting}
\end{figure}

Because of the short interaction range of C$_{60}$ molecules, and supported by the results of Fig.~\ref{fig:AllEnergyBarriers}, we decided to use the simplest version, which only considers the lateral nearest neighbors of the initial configuration. How the short interaction range of the Girifalco potential justifies this approach is illustrated in Fig.~\ref{fig:BondCounting}. As an example, starting from the three neighbor (3N) initial state $S_2$, the energy barriers for transition to the 3N final state $S_1$ or the 2N final state $S_3$ only have a negligible difference of $\SI{13}{\milli\electronvolt}$.

\section{Mean cluster curvature}
\label{sec:curv}

To calculate $\kappa$, we use the formula for the curvature of a continuous plane curve given in parametric form $\gamma(t)=(x(t),y(t))$,
\begin{equation}
  \kappa(t)=\frac{|\dot{x}(t)\ddot{y}(t)-\ddot{x}(t)\dot{y}(t)|}{(\dot{x}(t)^2+\dot{y}(t)^2)^\frac{3}{2}}.
  \label{eq:kappaCont}
\end{equation}
As the cluster border is not given as a continuous curve, but by a set of $N_E$ discrete border molecule positions, we approximate the derivatives at molecule $n$ as finite differences
\begin{gather}
  \dot{x}(n) = \frac{x(n+1)-x(n-1)}{2}, \\
  \ddot{x}(n) = x(n+1) + x(n-1) - 2x(n)
\end{gather}
($y$ coordinate analogous). Plugging these derivatives into Eq.~\eqref{eq:kappaCont}, we can calculate the mean border curvature via
\begin{equation}
  \bar\kappa =\frac{1}{N_E} \sum_{n=1}^{N_E}\kappa(n).
\end{equation}

%


\begin{thebibliography}{34}%
  \makeatletter
  \providecommand \@ifxundefined [1]{%
   \@ifx{#1\undefined}
  }%
  \providecommand \@ifnum [1]{%
   \ifnum #1\expandafter \@firstoftwo
   \else \expandafter \@secondoftwo
   \fi
  }%
  \providecommand \@ifx [1]{%
   \ifx #1\expandafter \@firstoftwo
   \else \expandafter \@secondoftwo
   \fi
  }%
  \providecommand \natexlab [1]{#1}%
  \providecommand \enquote  [1]{``#1''}%
  \providecommand \bibnamefont  [1]{#1}%
  \providecommand \bibfnamefont [1]{#1}%
  \providecommand \citenamefont [1]{#1}%
  \providecommand \href@noop [0]{\@secondoftwo}%
  \providecommand \href [0]{\begingroup \@sanitize@url \@href}%
  \providecommand \@href[1]{\@@startlink{#1}\@@href}%
  \providecommand \@@href[1]{\endgroup#1\@@endlink}%
  \providecommand \@sanitize@url [0]{\catcode `\\12\catcode `\$12\catcode
    `\&12\catcode `\#12\catcode `\^12\catcode `\_12\catcode `\%12\relax}%
  \providecommand \@@startlink[1]{}%
  \providecommand \@@endlink[0]{}%
  \providecommand \url  [0]{\begingroup\@sanitize@url \@url }%
  \providecommand \@url [1]{\endgroup\@href {#1}{\urlprefix }}%
  \providecommand \urlprefix  [0]{URL }%
  \providecommand \Eprint [0]{\href }%
  \providecommand \doibase [0]{http://dx.doi.org/}%
  \providecommand \selectlanguage [0]{\@gobble}%
  \providecommand \bibinfo  [0]{\@secondoftwo}%
  \providecommand \bibfield  [0]{\@secondoftwo}%
  \providecommand \translation [1]{[#1]}%
  \providecommand \BibitemOpen [0]{}%
  \providecommand \bibitemStop [0]{}%
  \providecommand \bibitemNoStop [0]{.\EOS\space}%
  \providecommand \EOS [0]{\spacefactor3000\relax}%
  \providecommand \BibitemShut  [1]{\csname bibitem#1\endcsname}%
  \let\auto@bib@innerbib\@empty
  \bibitem [{\citenamefont {Flood}\ \emph {et~al.}(2004)\citenamefont {Flood},
    \citenamefont {Stoddart}, \citenamefont {Steuerman},\ and\ \citenamefont
    {Heath}}]{Flood04}%
    \BibitemOpen
    \bibfield  {author} {\bibinfo {author} {\bibfnamefont {A.~H.}\ \bibnamefont
    {Flood}}, \bibinfo {author} {\bibfnamefont {J.~F.}\ \bibnamefont {Stoddart}},
    \bibinfo {author} {\bibfnamefont {D.~W.}\ \bibnamefont {Steuerman}}, \ and\
    \bibinfo {author} {\bibfnamefont {J.~R.}\ \bibnamefont {Heath}},\ }\bibfield
    {title} {\enquote {\bibinfo {title} {Whence molecular electronics?}}\ }\href
    {\doibase 10.1126/science.1106195} {\bibfield  {journal} {\bibinfo  {journal}
    {Science}\ }\textbf {\bibinfo {volume} {306}},\ \bibinfo {pages} {2055--2056}
    (\bibinfo {year} {2004})}\BibitemShut {NoStop}%
  \bibitem [{\citenamefont {Schulz}(1999)}]{SchulzEndofSilc}%
    \BibitemOpen
    \bibfield  {author} {\bibinfo {author} {\bibfnamefont {M.}~\bibnamefont
    {Schulz}},\ }\bibfield  {title} {\enquote {\bibinfo {title} {The end of the
    road for silicon?}}\ }\href {\doibase 10.1038/21526} {\bibfield  {journal}
    {\bibinfo  {journal} {Nature}\ }\textbf {\bibinfo {volume} {399}},\ \bibinfo
    {pages} {729--730} (\bibinfo {year} {1999})}\BibitemShut {NoStop}%
  \bibitem [{\citenamefont {Ning}(2010)}]{Ning2010SiliCMOSTech}%
    \BibitemOpen
    \bibfield  {author} {\bibinfo {author} {\bibfnamefont {T.~H.}\ \bibnamefont
    {Ning}},\ }\bibfield  {title} {\enquote {\bibinfo {title} {Directions for
    silicon technology as we approach the end of {CMOS} scaling},}\ }in\ \href
    {\doibase 10.1109/ICSICT.2010.5667878} {\emph {\bibinfo {booktitle} {2010
    10th IEEE International Conference on Solid-State and Integrated Circuit
    Technology}}}\ (\bibinfo {year} {2010})\ pp.\ \bibinfo {pages}
    {3--3}\BibitemShut {NoStop}%
  \bibitem [{\citenamefont {Graetzel}\ \emph {et~al.}(2012)\citenamefont
    {Graetzel}, \citenamefont {Janssen}, \citenamefont {Mitzi},\ and\
    \citenamefont {Sargent}}]{graetzel12}%
    \BibitemOpen
    \bibfield  {author} {\bibinfo {author} {\bibfnamefont {M.}~\bibnamefont
    {Graetzel}}, \bibinfo {author} {\bibfnamefont {R.~A.~J.}\ \bibnamefont
    {Janssen}}, \bibinfo {author} {\bibfnamefont {D.~B.}\ \bibnamefont {Mitzi}},
    \ and\ \bibinfo {author} {\bibfnamefont {E.~H.}\ \bibnamefont {Sargent}},\
    }\bibfield  {title} {\enquote {\bibinfo {title} {Materials interface
    engineering for solution-processed photovoltaics},}\ }\href {\doibase
    10.1038/nature11476} {\bibfield  {journal} {\bibinfo  {journal} {Nature}\
    }\textbf {\bibinfo {volume} {488}},\ \bibinfo {pages} {304--312} (\bibinfo
    {year} {2012})}\BibitemShut {NoStop}%
  \bibitem [{\citenamefont {Barth}\ \emph {et~al.}(2005)\citenamefont {Barth},
    \citenamefont {Costantini},\ and\ \citenamefont {Kern}}]{Barth05}%
    \BibitemOpen
    \bibfield  {author} {\bibinfo {author} {\bibfnamefont {J.~V.}\ \bibnamefont
    {Barth}}, \bibinfo {author} {\bibfnamefont {G.}~\bibnamefont {Costantini}}, \
    and\ \bibinfo {author} {\bibfnamefont {K.}~\bibnamefont {Kern}},\ }\bibfield
    {title} {\enquote {\bibinfo {title} {Engineering atomic and molecular
    nanostructures at surfaces},}\ }\href {\doibase 10.1038/nature04166}
    {\bibfield  {journal} {\bibinfo  {journal} {Nature}\ }\textbf {\bibinfo
    {volume} {437}},\ \bibinfo {pages} {671--679} (\bibinfo {year}
    {2005})}\BibitemShut {NoStop}%
  \bibitem [{\citenamefont {Barth}(2007)}]{Barth07}%
    \BibitemOpen
    \bibfield  {author} {\bibinfo {author} {\bibfnamefont {J.~V.}\ \bibnamefont
    {Barth}},\ }\bibfield  {title} {\enquote {\bibinfo {title} {Molecular
    architectonic on metal surfaces},}\ }\href {\doibase
    10.1146/annurev.physchem.56.092503.141259} {\bibfield  {journal} {\bibinfo
    {journal} {Annu. Rev. Phys. Chem.}\ }\textbf {\bibinfo {volume} {58}},\
    \bibinfo {pages} {375--407} (\bibinfo {year} {2007})}\BibitemShut {NoStop}%
  \bibitem [{\citenamefont {Kühnle}(2009)}]{kuhnle09}%
    \BibitemOpen
    \bibfield  {author} {\bibinfo {author} {\bibfnamefont {A.}~\bibnamefont
    {Kühnle}},\ }\bibfield  {title} {\enquote {\bibinfo {title} {Self-assembly
    of organic molecules at metal surfaces},}\ }\href {\doibase
    10.1016/j.cocis.2008.01.001} {\bibfield  {journal} {\bibinfo  {journal}
    {Curr. Opin. Colloid Interface Sci.}\ }\textbf {\bibinfo {volume} {14}},\
    \bibinfo {pages} {157--168} (\bibinfo {year} {2009})}\BibitemShut {NoStop}%
  \bibitem [{\citenamefont {van~der Lit}\ \emph {et~al.}(2016)\citenamefont
    {van~der Lit}, \citenamefont {Marsman}, \citenamefont {Koster}, \citenamefont
    {Jacobse}, \citenamefont {den Hartog}, \citenamefont {Vanmaekelbergh},
    \citenamefont {Gebbink}, \citenamefont {Filion},\ and\ \citenamefont
    {Swart}}]{lit15}%
    \BibitemOpen
    \bibfield  {author} {\bibinfo {author} {\bibfnamefont {J.}~\bibnamefont
    {van~der Lit}}, \bibinfo {author} {\bibfnamefont {J.~L.}\ \bibnamefont
    {Marsman}}, \bibinfo {author} {\bibfnamefont {R.~S.}\ \bibnamefont {Koster}},
    \bibinfo {author} {\bibfnamefont {P.~H.}\ \bibnamefont {Jacobse}}, \bibinfo
    {author} {\bibfnamefont {S.~A.}\ \bibnamefont {den Hartog}}, \bibinfo
    {author} {\bibfnamefont {D.}~\bibnamefont {Vanmaekelbergh}}, \bibinfo
    {author} {\bibfnamefont {R.~J. M.~K.}\ \bibnamefont {Gebbink}}, \bibinfo
    {author} {\bibfnamefont {L.}~\bibnamefont {Filion}}, \ and\ \bibinfo {author}
    {\bibfnamefont {I.}~\bibnamefont {Swart}},\ }\bibfield  {title} {\enquote
    {\bibinfo {title} {Modeling the self-assembly of organic molecules in {2D}
    molecular layers with different structures},}\ }\href {\doibase
    10.1021/acs.jpcc.5b09889} {\bibfield  {journal} {\bibinfo  {journal} {J.
    Phys. Chem. C}\ }\textbf {\bibinfo {volume} {120}},\ \bibinfo {pages}
    {318--323} (\bibinfo {year} {2016})}\BibitemShut {NoStop}%
  \bibitem [{\citenamefont {Liu}\ and\ \citenamefont {Reinke}(2006)}]{LiuC60Exp}%
    \BibitemOpen
    \bibfield  {author} {\bibinfo {author} {\bibfnamefont {H.}~\bibnamefont
    {Liu}}\ and\ \bibinfo {author} {\bibfnamefont {P.}~\bibnamefont {Reinke}},\
    }\bibfield  {title} {\enquote {\bibinfo {title} {{C}$_{60}$ thin film growth
    on graphite: Coexistence of spherical and fractal-dendritic islands},}\
    }\href {\doibase 10.1063/1.2186310} {\bibfield  {journal} {\bibinfo
    {journal} {J. Chem. Phys.}\ }\textbf {\bibinfo {volume} {124}},\ \bibinfo
    {pages} {164707} (\bibinfo {year} {2006})}\BibitemShut {NoStop}%
  \bibitem [{\citenamefont {Conrad}\ \emph {et~al.}(2009)\citenamefont {Conrad},
    \citenamefont {Tosado}, \citenamefont {Dutton}, \citenamefont {Dougherty},
    \citenamefont {Jin}, \citenamefont {Bonnen}, \citenamefont {Schuldenfrei},
    \citenamefont {Cullen}, \citenamefont {Williams}, \citenamefont
    {Reutt-Robey},\ and\ \citenamefont {Robey}}]{conrad09}%
    \BibitemOpen
    \bibfield  {author} {\bibinfo {author} {\bibfnamefont {B.~R.}\ \bibnamefont
    {Conrad}}, \bibinfo {author} {\bibfnamefont {J.}~\bibnamefont {Tosado}},
    \bibinfo {author} {\bibfnamefont {G.}~\bibnamefont {Dutton}}, \bibinfo
    {author} {\bibfnamefont {D.~B.}\ \bibnamefont {Dougherty}}, \bibinfo {author}
    {\bibfnamefont {W.}~\bibnamefont {Jin}}, \bibinfo {author} {\bibfnamefont
    {T.}~\bibnamefont {Bonnen}}, \bibinfo {author} {\bibfnamefont
    {A.}~\bibnamefont {Schuldenfrei}}, \bibinfo {author} {\bibfnamefont {W.~G.}\
    \bibnamefont {Cullen}}, \bibinfo {author} {\bibfnamefont {E.~D.}\
    \bibnamefont {Williams}}, \bibinfo {author} {\bibfnamefont {J.~E.}\
    \bibnamefont {Reutt-Robey}}, \ and\ \bibinfo {author} {\bibfnamefont {S.~W.}\
    \bibnamefont {Robey}},\ }\bibfield  {title} {\enquote {\bibinfo {title}
    {C$_{60}$ cluster formation at interfaces with pentacene thin-film phases},}\
    }\href {\doibase 10.1063/1.3266857} {\bibfield  {journal} {\bibinfo
    {journal} {Appl. Phys. Lett.}\ }\textbf {\bibinfo {volume} {95}},\ \bibinfo
    {pages} {213302} (\bibinfo {year} {2009})}\BibitemShut {NoStop}%
  \bibitem [{\citenamefont {Loske}\ \emph {et~al.}(2010)\citenamefont {Loske},
    \citenamefont {L\"ubbe}, \citenamefont {Sch\"utte}, \citenamefont
    {Reichling},\ and\ \citenamefont {K\"uhnle}}]{LoskePaper}%
    \BibitemOpen
    \bibfield  {author} {\bibinfo {author} {\bibfnamefont {F.}~\bibnamefont
    {Loske}}, \bibinfo {author} {\bibfnamefont {J.}~\bibnamefont {L\"ubbe}},
    \bibinfo {author} {\bibfnamefont {J.}~\bibnamefont {Sch\"utte}}, \bibinfo
    {author} {\bibfnamefont {M.}~\bibnamefont {Reichling}}, \ and\ \bibinfo
    {author} {\bibfnamefont {A.}~\bibnamefont {K\"uhnle}},\ }\bibfield  {title}
    {\enquote {\bibinfo {title} {Quantitative description of {C}$_{60}$ diffusion
    on an insulating surface},}\ }\href {\doibase 10.1103/PhysRevB.82.155428}
    {\bibfield  {journal} {\bibinfo  {journal} {Phys. Rev. B}\ }\textbf {\bibinfo
    {volume} {82}},\ \bibinfo {pages} {155428} (\bibinfo {year}
    {2010})}\BibitemShut {NoStop}%
  \bibitem [{\citenamefont {K\"orner}\ \emph {et~al.}(2011)\citenamefont
    {K\"orner}, \citenamefont {Loske}, \citenamefont {Einax}, \citenamefont
    {K\"uhnle}, \citenamefont {Reichling},\ and\ \citenamefont {Maass}}]{Korn11}%
    \BibitemOpen
    \bibfield  {author} {\bibinfo {author} {\bibfnamefont {M.}~\bibnamefont
    {K\"orner}}, \bibinfo {author} {\bibfnamefont {F.}~\bibnamefont {Loske}},
    \bibinfo {author} {\bibfnamefont {M.}~\bibnamefont {Einax}}, \bibinfo
    {author} {\bibfnamefont {A.}~\bibnamefont {K\"uhnle}}, \bibinfo {author}
    {\bibfnamefont {M.}~\bibnamefont {Reichling}}, \ and\ \bibinfo {author}
    {\bibfnamefont {P.}~\bibnamefont {Maass}},\ }\bibfield  {title} {\enquote
    {\bibinfo {title} {Second-layer induced island morphologies in thin-film
    growth of fullerenes},}\ }\href {\doibase 10.1103/PhysRevLett.107.016101}
    {\bibfield  {journal} {\bibinfo  {journal} {Phys. Rev. Lett.}\ }\textbf
    {\bibinfo {volume} {107}},\ \bibinfo {pages} {016101} (\bibinfo {year}
    {2011})}\BibitemShut {NoStop}%
  \bibitem [{\citenamefont {Bommel}\ \emph {et~al.}(2014)\citenamefont {Bommel},
    \citenamefont {Kleppmann}, \citenamefont {Weber}, \citenamefont {Spranger},
    \citenamefont {Sch\"afer}, \citenamefont {Novak}, \citenamefont {Roth},
    \citenamefont {Schreiber}, \citenamefont {Klapp},\ and\ \citenamefont
    {Kowarik}}]{bomm14}%
    \BibitemOpen
    \bibfield  {author} {\bibinfo {author} {\bibfnamefont {S.}~\bibnamefont
    {Bommel}}, \bibinfo {author} {\bibfnamefont {N.}~\bibnamefont {Kleppmann}},
    \bibinfo {author} {\bibfnamefont {C.}~\bibnamefont {Weber}}, \bibinfo
    {author} {\bibfnamefont {H.}~\bibnamefont {Spranger}}, \bibinfo {author}
    {\bibfnamefont {P.}~\bibnamefont {Sch\"afer}}, \bibinfo {author}
    {\bibfnamefont {J.}~\bibnamefont {Novak}}, \bibinfo {author} {\bibfnamefont
    {S.}~\bibnamefont {Roth}}, \bibinfo {author} {\bibfnamefont {F.}~\bibnamefont
    {Schreiber}}, \bibinfo {author} {\bibfnamefont {S.}~\bibnamefont {Klapp}}, \
    and\ \bibinfo {author} {\bibfnamefont {S.}~\bibnamefont {Kowarik}},\
    }\bibfield  {title} {\enquote {\bibinfo {title} {Unravelling the multilayer
    growth of the fullerene {C}$_{60}$ in real time},}\ }\href {\doibase
    10.1038/ncomms6388} {\bibfield  {journal} {\bibinfo  {journal} {Nat.
    Commun.}\ }\textbf {\bibinfo {volume} {5}},\ \bibinfo {pages} {5388}
    (\bibinfo {year} {2014})}\BibitemShut {NoStop}%
  \bibitem [{\citenamefont {Picone}\ \emph {et~al.}(2016)\citenamefont {Picone},
    \citenamefont {Giannotti}, \citenamefont {Riva}, \citenamefont {Calloni},
    \citenamefont {Bussetti}, \citenamefont {Berti}, \citenamefont {Duò},
    \citenamefont {Ciccacci}, \citenamefont {Finazzi},\ and\ \citenamefont
    {Brambilla}}]{Picone16}%
    \BibitemOpen
    \bibfield  {author} {\bibinfo {author} {\bibfnamefont {A.}~\bibnamefont
    {Picone}}, \bibinfo {author} {\bibfnamefont {D.}~\bibnamefont {Giannotti}},
    \bibinfo {author} {\bibfnamefont {M.}~\bibnamefont {Riva}}, \bibinfo {author}
    {\bibfnamefont {A.}~\bibnamefont {Calloni}}, \bibinfo {author} {\bibfnamefont
    {G.}~\bibnamefont {Bussetti}}, \bibinfo {author} {\bibfnamefont
    {G.}~\bibnamefont {Berti}}, \bibinfo {author} {\bibfnamefont
    {L.}~\bibnamefont {Duò}}, \bibinfo {author} {\bibfnamefont {F.}~\bibnamefont
    {Ciccacci}}, \bibinfo {author} {\bibfnamefont {M.}~\bibnamefont {Finazzi}}, \
    and\ \bibinfo {author} {\bibfnamefont {A.}~\bibnamefont {Brambilla}},\
    }\bibfield  {title} {\enquote {\bibinfo {title} {Controlling the electronic
    and structural coupling of {C}$_{60}$ nano films on {F}e(001) through oxygen
    adsorption at the interface},}\ }\href {\doibase 10.1021/acsami.6b09641}
    {\bibfield  {journal} {\bibinfo  {journal} {ACS Applied Materials \&
    Interfaces}\ }\textbf {\bibinfo {volume} {8}},\ \bibinfo {pages}
    {26418--26424} (\bibinfo {year} {2016})}\BibitemShut {NoStop}%
  \bibitem [{\citenamefont {Einax}\ \emph {et~al.}(2013)\citenamefont {Einax},
    \citenamefont {Dieterich},\ and\ \citenamefont {Maass}}]{einax13}%
    \BibitemOpen
    \bibfield  {author} {\bibinfo {author} {\bibfnamefont {M.}~\bibnamefont
    {Einax}}, \bibinfo {author} {\bibfnamefont {W.}~\bibnamefont {Dieterich}}, \
    and\ \bibinfo {author} {\bibfnamefont {P.}~\bibnamefont {Maass}},\ }\bibfield
     {title} {\enquote {\bibinfo {title} {Colloquium: Cluster growth on surfaces:
    Densities, size distributions, and morphologies},}\ }\href {\doibase
    10.1103/RevModPhys.85.921} {\bibfield  {journal} {\bibinfo  {journal} {Rev.
    Mod. Phys.}\ }\textbf {\bibinfo {volume} {85}},\ \bibinfo {pages} {921--939}
    (\bibinfo {year} {2013})}\BibitemShut {NoStop}%
  \bibitem [{\citenamefont {Voter}(2007)}]{KMCVoter2007}%
    \BibitemOpen
    \bibfield  {author} {\bibinfo {author} {\bibfnamefont {A.~F.}\ \bibnamefont
    {Voter}},\ }\enquote {\bibinfo {title}
    {\href{http://link.springer.com/chapter/10.1007/978-1-4020-5295-8_1}{Introduction
    to the Kinetic Monte Carlo Method}},}\ in\ \href {\doibase
    10.1007/978-1-4020-5295-8_1} {\emph {\bibinfo {booktitle} {Radiation Effects
    in Solids}}},\ \bibinfo {editor} {edited by\ \bibinfo {editor} {\bibfnamefont
    {K.~E.}\ \bibnamefont {Sickafus}}, \bibinfo {editor} {\bibfnamefont {E.~A.}\
    \bibnamefont {Kotomin}}, \ and\ \bibinfo {editor} {\bibfnamefont {B.~P.}\
    \bibnamefont {Uberuaga}}}\ (\bibinfo  {publisher} {Springer Netherlands},\
    \bibinfo {address} {Dordrecht},\ \bibinfo {year} {2007})\ pp.\ \bibinfo
    {pages} {1--23}\BibitemShut {NoStop}%
  \bibitem [{\citenamefont {Gillespie}(1976)}]{Gillespie76}%
    \BibitemOpen
    \bibfield  {author} {\bibinfo {author} {\bibfnamefont {D.~T.}\ \bibnamefont
    {Gillespie}},\ }\bibfield  {title} {\enquote {\bibinfo {title} {A general
    method for numerically simulating the stochastic time evolution of coupled
    chemical reactions},}\ }\href {\doibase 10.1016/0021-9991(76)90041-3}
    {\bibfield  {journal} {\bibinfo  {journal} {J. Comput. Phys.}\ }\textbf
    {\bibinfo {volume} {22}},\ \bibinfo {pages} {403--434} (\bibinfo {year}
    {1976})}\BibitemShut {NoStop}%
  \bibitem [{\citenamefont {Gillespie}(1977)}]{Gillespie77}%
    \BibitemOpen
    \bibfield  {author} {\bibinfo {author} {\bibfnamefont {D.~T.}\ \bibnamefont
    {Gillespie}},\ }\bibfield  {title} {\enquote {\bibinfo {title} {Exact
    stochastic simulation of coupled chemical reactions},}\ }\href {\doibase
    10.1021/j100540a008} {\bibfield  {journal} {\bibinfo  {journal} {J. Phys.
    Chem.}\ }\textbf {\bibinfo {volume} {81}},\ \bibinfo {pages} {2340--2361}
    (\bibinfo {year} {1977})}\BibitemShut {NoStop}%
  \bibitem [{\citenamefont {Liu}\ \emph {et~al.}(2008)\citenamefont {Liu},
    \citenamefont {Lin}, \citenamefont {Zhigilei},\ and\ \citenamefont
    {Reinke}}]{LiuC60Sim}%
    \BibitemOpen
    \bibfield  {author} {\bibinfo {author} {\bibfnamefont {H.}~\bibnamefont
    {Liu}}, \bibinfo {author} {\bibfnamefont {Z.}~\bibnamefont {Lin}}, \bibinfo
    {author} {\bibfnamefont {L.~V.}\ \bibnamefont {Zhigilei}}, \ and\ \bibinfo
    {author} {\bibfnamefont {P.}~\bibnamefont {Reinke}},\ }\bibfield  {title}
    {\enquote {\bibinfo {title} {Fractal structures in fullerene layers:
    Simulation of the growth process},}\ }\href {\doibase 10.1021/jp0775597}
    {\bibfield  {journal} {\bibinfo  {journal} {J. Phys. Chem. C}\ }\textbf
    {\bibinfo {volume} {112}},\ \bibinfo {pages} {4687--4695} (\bibinfo {year}
    {2008})}\BibitemShut {NoStop}%
  \bibitem [{\citenamefont {Kleppmann}\ and\ \citenamefont
    {Klapp}(2015)}]{Kleppmann15}%
    \BibitemOpen
    \bibfield  {author} {\bibinfo {author} {\bibfnamefont {N.}~\bibnamefont
    {Kleppmann}}\ and\ \bibinfo {author} {\bibfnamefont {S.~H.~L.}\ \bibnamefont
    {Klapp}},\ }\bibfield  {title} {\enquote {\bibinfo {title} {Particle-resolved
    dynamics during multilayer growth of {C}$_{60}$},}\ }\href {\doibase
    10.1103/PhysRevB.91.045436} {\bibfield  {journal} {\bibinfo  {journal} {Phys.
    Rev. B}\ }\textbf {\bibinfo {volume} {91}},\ \bibinfo {pages} {045436}
    (\bibinfo {year} {2015})}\BibitemShut {NoStop}%
  \bibitem [{\citenamefont {Acevedo}\ \emph {et~al.}(2016)\citenamefont
    {Acevedo}, \citenamefont {Cantrell}, \citenamefont {Berard}, \citenamefont
    {Koch},\ and\ \citenamefont {Clancy}}]{acevedo16}%
    \BibitemOpen
    \bibfield  {author} {\bibinfo {author} {\bibfnamefont {Y.~M.}\ \bibnamefont
    {Acevedo}}, \bibinfo {author} {\bibfnamefont {R.~A.}\ \bibnamefont
    {Cantrell}}, \bibinfo {author} {\bibfnamefont {P.~G.}\ \bibnamefont
    {Berard}}, \bibinfo {author} {\bibfnamefont {D.~L.}\ \bibnamefont {Koch}}, \
    and\ \bibinfo {author} {\bibfnamefont {P.}~\bibnamefont {Clancy}},\
    }\bibfield  {title} {\enquote {\bibinfo {title} {Multiscale simulation and
    modeling of multilayer heteroepitactic growth of {C}$_{60}$ on pentacene},}\
    }\href {\doibase 10.1021/acs.langmuir.5b04500} {\bibfield  {journal}
    {\bibinfo  {journal} {Langmuir}\ }\textbf {\bibinfo {volume} {32}},\ \bibinfo
    {pages} {3045--3056} (\bibinfo {year} {2016})}\BibitemShut {NoStop}%
  \bibitem [{\citenamefont {Kleppmann}\ \emph {et~al.}(2017)\citenamefont
    {Kleppmann}, \citenamefont {Schreiber},\ and\ \citenamefont
    {Klapp}}]{Kleppmann17}%
    \BibitemOpen
    \bibfield  {author} {\bibinfo {author} {\bibfnamefont {N.}~\bibnamefont
    {Kleppmann}}, \bibinfo {author} {\bibfnamefont {F.}~\bibnamefont
    {Schreiber}}, \ and\ \bibinfo {author} {\bibfnamefont {S.~H.~L.}\
    \bibnamefont {Klapp}},\ }\bibfield  {title} {\enquote {\bibinfo {title}
    {Limits of size scalability of diffusion and growth: Atoms versus molecules
    versus colloids},}\ }\href {\doibase 10.1103/PhysRevE.95.020801} {\bibfield
    {journal} {\bibinfo  {journal} {Phys. Rev. E}\ }\textbf {\bibinfo {volume}
    {95}},\ \bibinfo {pages} {020801} (\bibinfo {year} {2017})}\BibitemShut
    {NoStop}%
  \bibitem [{\citenamefont {Goose}\ \emph {et~al.}(2010)\citenamefont {Goose},
    \citenamefont {First},\ and\ \citenamefont {Clancy}}]{GooseDFTDiffCalc}%
    \BibitemOpen
    \bibfield  {author} {\bibinfo {author} {\bibfnamefont {J.~E.}\ \bibnamefont
    {Goose}}, \bibinfo {author} {\bibfnamefont {E.~L.}\ \bibnamefont {First}}, \
    and\ \bibinfo {author} {\bibfnamefont {P.}~\bibnamefont {Clancy}},\
    }\bibfield  {title} {\enquote {\bibinfo {title} {Nature of step-edge barriers
    for small organic molecules},}\ }\href {\doibase 10.1103/PhysRevB.81.205310}
    {\bibfield  {journal} {\bibinfo  {journal} {Phys. Rev. B}\ }\textbf {\bibinfo
    {volume} {81}},\ \bibinfo {pages} {205310} (\bibinfo {year}
    {2010})}\BibitemShut {NoStop}%
  \bibitem [{\citenamefont {Cantrell}\ and\ \citenamefont
    {Clancy}(2012)}]{Cantrell12}%
    \BibitemOpen
    \bibfield  {author} {\bibinfo {author} {\bibfnamefont {R.~A.}\ \bibnamefont
    {Cantrell}}\ and\ \bibinfo {author} {\bibfnamefont {P.}~\bibnamefont
    {Clancy}},\ }\bibfield  {title} {\enquote {\bibinfo {title} {A new kinetic
    monte carlo algorithm for heteroepitactical growth: Case study of {C}$_{60}$
    growth on pentacene},}\ }\href {\doibase 10.1021/ct200819r} {\bibfield
    {journal} {\bibinfo  {journal} {J. Chem. Theory Comput.}\ }\textbf {\bibinfo
    {volume} {8}},\ \bibinfo {pages} {1048--1057} (\bibinfo {year}
    {2012})}\BibitemShut {NoStop}%
  \bibitem [{\citenamefont {Plimpton}(1995)}]{LAMMPSPlimpton}%
    \BibitemOpen
    \bibfield  {author} {\bibinfo {author} {\bibfnamefont {S.}~\bibnamefont
    {Plimpton}},\ }\bibfield  {title} {\enquote {\bibinfo {title} {Fast parallel
    algorithms for short-range molecular dynamics},}\ }\href {\doibase
    http://dx.doi.org/10.1006/jcph.1995.1039} {\bibfield  {journal} {\bibinfo
    {journal} {J. Comput. Phys.}\ }\textbf {\bibinfo {volume} {117}},\ \bibinfo
    {pages} {1 -- 19} (\bibinfo {year} {1995})}\BibitemShut {NoStop}%
  \bibitem [{\citenamefont {Girifalco}(1992)}]{GirifalcoPot92}%
    \BibitemOpen
    \bibfield  {author} {\bibinfo {author} {\bibfnamefont {L.~A.}\ \bibnamefont
    {Girifalco}},\ }\bibfield  {title} {\enquote {\bibinfo {title} {Molecular
    properties of fullerene in the gas and solid phases},}\ }\href {\doibase
    10.1021/j100181a061} {\bibfield  {journal} {\bibinfo  {journal} {J. Phys.
    Chem.}\ }\textbf {\bibinfo {volume} {96}},\ \bibinfo {pages} {858--861}
    (\bibinfo {year} {1992})}\BibitemShut {NoStop}%
  \bibitem [{\citenamefont {Girifalco}(1991)}]{GirifalcoPot}%
    \BibitemOpen
    \bibfield  {author} {\bibinfo {author} {\bibfnamefont {L.~A.}\ \bibnamefont
    {Girifalco}},\ }\bibfield  {title} {\enquote {\bibinfo {title} {Interaction
    potential for {C}$_{60}$ molecules},}\ }\href {\doibase 10.1021/j100167a002}
    {\bibfield  {journal} {\bibinfo  {journal} {J. Phys. Chem.}\ }\textbf
    {\bibinfo {volume} {95}},\ \bibinfo {pages} {5370--5371} (\bibinfo {year}
    {1991})}\BibitemShut {NoStop}%
  \bibitem [{\citenamefont {Chiutu}\ \emph {et~al.}(2012)\citenamefont {Chiutu},
    \citenamefont {Sweetman}, \citenamefont {Lakin}, \citenamefont {Stannard},
    \citenamefont {Jarvis}, \citenamefont {Kantorovich}, \citenamefont {Dunn},\
    and\ \citenamefont {Moriarty}}]{ChiutuPotMeasure}%
    \BibitemOpen
    \bibfield  {author} {\bibinfo {author} {\bibfnamefont {C.}~\bibnamefont
    {Chiutu}}, \bibinfo {author} {\bibfnamefont {A.~M.}\ \bibnamefont
    {Sweetman}}, \bibinfo {author} {\bibfnamefont {A.~J.}\ \bibnamefont {Lakin}},
    \bibinfo {author} {\bibfnamefont {A.}~\bibnamefont {Stannard}}, \bibinfo
    {author} {\bibfnamefont {S.}~\bibnamefont {Jarvis}}, \bibinfo {author}
    {\bibfnamefont {L.}~\bibnamefont {Kantorovich}}, \bibinfo {author}
    {\bibfnamefont {J.~L.}\ \bibnamefont {Dunn}}, \ and\ \bibinfo {author}
    {\bibfnamefont {P.}~\bibnamefont {Moriarty}},\ }\bibfield  {title} {\enquote
    {\bibinfo {title} {Precise orientation of a single {C}$_{60}$ molecule on the
    tip of a scanning probe microscope},}\ }\href {\doibase
    10.1103/PhysRevLett.108.268302} {\bibfield  {journal} {\bibinfo  {journal}
    {Phys. Rev. Lett.}\ }\textbf {\bibinfo {volume} {108}},\ \bibinfo {pages}
    {268302} (\bibinfo {year} {2012})}\BibitemShut {NoStop}%
  \bibitem [{\citenamefont {Khlobystov}\ \emph {et~al.}(2004)\citenamefont
    {Khlobystov}, \citenamefont {Britz}, \citenamefont {Ardavan},\ and\
    \citenamefont {Briggs}}]{khlob04}%
    \BibitemOpen
    \bibfield  {author} {\bibinfo {author} {\bibfnamefont {A.~N.}\ \bibnamefont
    {Khlobystov}}, \bibinfo {author} {\bibfnamefont {D.~A.}\ \bibnamefont
    {Britz}}, \bibinfo {author} {\bibfnamefont {A.}~\bibnamefont {Ardavan}}, \
    and\ \bibinfo {author} {\bibfnamefont {G.~A.~D.}\ \bibnamefont {Briggs}},\
    }\bibfield  {title} {\enquote {\bibinfo {title} {Observation of ordered
    phases of fullerenes in carbon nanotubes},}\ }\href {\doibase
    10.1103/PhysRevLett.92.245507} {\bibfield  {journal} {\bibinfo  {journal}
    {Phys. Rev. Lett.}\ }\textbf {\bibinfo {volume} {92}},\ \bibinfo {pages}
    {245507} (\bibinfo {year} {2004})}\BibitemShut {NoStop}%
  \bibitem [{\citenamefont {Gravil}\ \emph {et~al.}(1996)\citenamefont {Gravil},
    \citenamefont {Devel}, \citenamefont {Lambin}, \citenamefont {Bouju},
    \citenamefont {Girard},\ and\ \citenamefont {Lucas}}]{GravilPPotDiffCalc}%
    \BibitemOpen
    \bibfield  {author} {\bibinfo {author} {\bibfnamefont {P.~A.}\ \bibnamefont
    {Gravil}}, \bibinfo {author} {\bibfnamefont {M.}~\bibnamefont {Devel}},
    \bibinfo {author} {\bibfnamefont {P.}~\bibnamefont {Lambin}}, \bibinfo
    {author} {\bibfnamefont {X.}~\bibnamefont {Bouju}}, \bibinfo {author}
    {\bibfnamefont {C.}~\bibnamefont {Girard}}, \ and\ \bibinfo {author}
    {\bibfnamefont {A.~A.}\ \bibnamefont {Lucas}},\ }\bibfield  {title} {\enquote
    {\bibinfo {title} {Adsorption of {C}$_{60}$ molecules},}\ }\href {\doibase
    10.1103/PhysRevB.53.1622} {\bibfield  {journal} {\bibinfo  {journal} {Phys.
    Rev. B}\ }\textbf {\bibinfo {volume} {53}},\ \bibinfo {pages} {1622--1629}
    (\bibinfo {year} {1996})}\BibitemShut {NoStop}%
  \bibitem [{\citenamefont {Seifert}(2005)}]{Seifert2005EntropyProd}%
    \BibitemOpen
    \bibfield  {author} {\bibinfo {author} {\bibfnamefont {U.}~\bibnamefont
    {Seifert}},\ }\bibfield  {title} {\enquote {\bibinfo {title} {Entropy
    production along a stochastic trajectory and an integral fluctuation
    theorem},}\ }\href {\doibase 10.1103/PhysRevLett.95.040602} {\bibfield
    {journal} {\bibinfo  {journal} {Phys. Rev. Lett.}\ }\textbf {\bibinfo
    {volume} {95}},\ \bibinfo {pages} {040602} (\bibinfo {year}
    {2005})}\BibitemShut {NoStop}%
  \bibitem [{\citenamefont {Renaud}\ \emph {et~al.}(2003)\citenamefont {Renaud},
    \citenamefont {Lazzari}, \citenamefont {Revenant}, \citenamefont {Barbier},
    \citenamefont {Noblet}, \citenamefont {Ulrich}, \citenamefont {Leroy},
    \citenamefont {Jupille}, \citenamefont {Borensztein}, \citenamefont {Henry},
    \citenamefont {Deville}, \citenamefont {Scheurer}, \citenamefont
    {Mane-Mane},\ and\ \citenamefont {Fruchart}}]{Renaud03}%
    \BibitemOpen
    \bibfield  {author} {\bibinfo {author} {\bibfnamefont {G.}~\bibnamefont
    {Renaud}}, \bibinfo {author} {\bibfnamefont {R.}~\bibnamefont {Lazzari}},
    \bibinfo {author} {\bibfnamefont {C.}~\bibnamefont {Revenant}}, \bibinfo
    {author} {\bibfnamefont {A.}~\bibnamefont {Barbier}}, \bibinfo {author}
    {\bibfnamefont {M.}~\bibnamefont {Noblet}}, \bibinfo {author} {\bibfnamefont
    {O.}~\bibnamefont {Ulrich}}, \bibinfo {author} {\bibfnamefont
    {F.}~\bibnamefont {Leroy}}, \bibinfo {author} {\bibfnamefont
    {J.}~\bibnamefont {Jupille}}, \bibinfo {author} {\bibfnamefont
    {Y.}~\bibnamefont {Borensztein}}, \bibinfo {author} {\bibfnamefont {C.~R.}\
    \bibnamefont {Henry}}, \bibinfo {author} {\bibfnamefont {J.-P.}\ \bibnamefont
    {Deville}}, \bibinfo {author} {\bibfnamefont {F.}~\bibnamefont {Scheurer}},
    \bibinfo {author} {\bibfnamefont {J.}~\bibnamefont {Mane-Mane}}, \ and\
    \bibinfo {author} {\bibfnamefont {O.}~\bibnamefont {Fruchart}},\ }\bibfield
    {title} {\enquote {\bibinfo {title} {Real-time monitoring of growing
    nanoparticles},}\ }\href {\doibase 10.1126/science.1082146} {\bibfield
    {journal} {\bibinfo  {journal} {Science}\ }\textbf {\bibinfo {volume}
    {300}},\ \bibinfo {pages} {1416--1419} (\bibinfo {year} {2003})}\BibitemShut
    {NoStop}%
  \bibitem [{\citenamefont {Bommel}(2015)}]{bomm15phd}%
    \BibitemOpen
    \bibfield  {author} {\bibinfo {author} {\bibfnamefont {S.}~\bibnamefont
    {Bommel}},\ }\emph {\bibinfo {title} {Unravelling nanoscale molecular
    processes in organic thin films}},\ \href {\doibase
    http://dx.doi.org/10.18452/17315} {Ph.D. thesis},\ \bibinfo  {school}
    {Humboldt-Universität zu Berlin, Mathematisch-Naturwissenschaftliche
    Fakultät} (\bibinfo {year} {2015})\BibitemShut {NoStop}%
  \bibitem [{\citenamefont {Rahe}\ \emph {et~al.}(2012)\citenamefont {Rahe},
    \citenamefont {Lindner}, \citenamefont {Kittelmann}, \citenamefont
    {Nimmrich},\ and\ \citenamefont {K\"uhnle}}]{rahe12}%
    \BibitemOpen
    \bibfield  {author} {\bibinfo {author} {\bibfnamefont {P.}~\bibnamefont
    {Rahe}}, \bibinfo {author} {\bibfnamefont {R.}~\bibnamefont {Lindner}},
    \bibinfo {author} {\bibfnamefont {M.}~\bibnamefont {Kittelmann}}, \bibinfo
    {author} {\bibfnamefont {M.}~\bibnamefont {Nimmrich}}, \ and\ \bibinfo
    {author} {\bibfnamefont {A.}~\bibnamefont {K\"uhnle}},\ }\bibfield  {title}
    {\enquote {\bibinfo {title} {From dewetting to wetting molecular layers:
    {C}$_{60}$ on {CaCO}$_3$(10$\bar 1$4) as a case study},}\ }\href {\doibase
    10.1039/C2CP40172J} {\bibfield  {journal} {\bibinfo  {journal} {Phys. Chem.
    Chem. Phys.}\ }\textbf {\bibinfo {volume} {14}},\ \bibinfo {pages}
    {6544--6548} (\bibinfo {year} {2012})}\BibitemShut {NoStop}%
  \end{thebibliography}
\end{document}